\documentclass[conference]{IEEEtran}
\IEEEoverridecommandlockouts

\newcommand\camerareadyversion[2]{#2}

\camerareadyversion{}{\pagestyle{plain}}
\usepackage{graphicx} 
\usepackage{mdframed}
\usepackage{algorithm}
\usepackage{microtype}
\usepackage{textcomp}
\usepackage[noend]{algpseudocode}
\usepackage[
 svgnames,
 table,
]{xcolor}

\usepackage[
 colorlinks,
 urlcolor = NavyBlue,
 citecolor = NavyBlue,
 linkcolor = NavyBlue,
 ]{hyperref}

\usepackage{mathtools}
\usepackage{amsthm}
\theoremstyle{plain}
\newtheorem{definition}{Definition}
\newtheorem{theorem}{Theorem}
\newtheorem*{theorem*}{Theorem}
\newtheorem{lemma}{Lemma}
\newtheorem*{lemma*}{Lemma}
\newtheorem{example}{Example}

\newtheorem*{corollary*}{Corollary}
\usepackage{biblatex} 
\usepackage{enumitem}
\newcommand{\myparagraph}[1]{\smallskip\noindent \textit{#1}}
\usepackage{xspace}
\usepackage{multirow}
\usepackage[pdf]{graphviz}
\usepackage{subcaption}

\usepackage{booktabs}
\usepackage{adjustbox}

\newenvironment{problemstatement}[1]
  {\mdfsetup{
    frametitle={\colorbox{white}{\space#1\space}},
    nobreak=true,
    innertopmargin=-4pt,
    frametitleaboveskip=-\ht\strutbox,
    frametitlealignment=\center,
    skipbelow=10pt
    }
  \begin{mdframed}%
  }
  {\end{mdframed}}
  
\usepackage{tikz}
\usetikzlibrary{automata,
                arrows.meta,
                chains,
                positioning,
                quotes,
                shapes.geometric,
                patterns,
                backgrounds}
\tikzset{
initial text=$ $,
treenode/.style = {align=center, inner sep=0pt, text centered},
  basis/.style = {
    pattern=north east lines,
    pattern color=magenta!70!black,
  },
basis/.style={
    pattern=north east lines,
    pattern color=black!40!white,
  },
}
\usepackage[outline]{contour}
\contourlength{1.2pt}
\newcommand{\treeNodeLabel}[1]{\contour{white}{#1}}

\usepackage[draft,silent]{fixme}
\fxsetup{theme=color,mode=multiuser}
\definecolor{fxtarget}{rgb}{0.8000,0.0000,0.0000}

\newcommand{\Specs}{\ensuremath{\mathbb{M}}\xspace}
\newcommand{\Impls}{\ensuremath{\mathbb{I}}\xspace}
\newcommand{\Spec}{\ensuremath{\mathcal{M}}\xspace}
\newcommand{\Impl}{\ensuremath{\mathcal{I}}\xspace}
\newcommand{\M}{\ensuremath{\mathcal{M}}\xspace}

\newcommand{\ifalg}{\textsc{Infernal}\xspace}
\newcommand{\lsharp}{\ensuremath{L^{\#}}\xspace}
\newcommand{\adaptivelsharp}{\ensuremath{AL^{\#}}\xspace}
\newcommand{\rlsharp}{\ensuremath{RL^{\#}}\xspace}
\newcommand{\radaptivelsharp}{\ensuremath{RAL^{\#}}\xspace}

\newcommand{\etal}{\emph{et al}.\xspace}
\usepackage{xcolor}
\definecolor{myorange}{HTML}{ff7f0e}
\definecolor{mygreen}{HTML}{2ca02c}
\definecolor{myblue}{HTML}{1f77b4}

\usepackage{tikz}
\usetikzlibrary{shapes.geometric, arrows}

\tikzstyle{startstop} = [rectangle, rounded corners, 
minimum width=3cm, 
minimum height=1cm,
text centered, 
draw=black, 
fill=mygreen!50]
\tikzstyle{io} = [rectangle, rounded corners, 
minimum width=13cm, 
minimum height=1cm,
text centered, 
draw=black, 
fill=myblue!50]
\tikzstyle{process} = [rectangle, 
minimum width=2cm, 
minimum height=1cm, 
text centered, 
text width=2cm, 
draw=black, 
fill=myorange!50]
\tikzstyle{background} = [rectangle, 
minimum width=16cm, 
minimum height=5cm, 
text centered, 
text width=3cm, 
draw=black, 
fill=black!15]
\tikzstyle{full_background} = [rectangle, 
minimum width=21cm, 
minimum height=6cm, 
text centered, 
text width=3cm, 
draw=black, 
xshift=-1.9cm,
fill=black!8]
\tikzstyle{arrow} = [thick,->,>=stealth]

\usepackage{amssymb}
\usepackage{amsmath}
\usepackage{enumitem}

\usepackage{makecell}

\usepackage{xspace}

\makeatletter
\newcommand{\linebreakand}{%
  \end{@IEEEauthorhalign}
  \hfill\mbox{}\par
  \mbox{}\hfill\begin{@IEEEauthorhalign}
}
\makeatother

\title{
Incremental Fingerprinting in an Open World}

\def\BibTeX{{\rm B\kern-.05em{\sc i\kern-.025em b}\kern-.08em
    T\kern-.1667em\lower.7ex\hbox{E}\kern-.125emX}}
\begin{document}

\title{Incremental Fingerprinting in an Open World\\
\thanks{This research is partially supported by the NWO grant No.~VI.Vidi.223.096.}
}
\author{%
\IEEEauthorblockN{Loes Kruger}
\IEEEauthorblockA{
\textit{Radboud University}\\
Nijmegen, The Netherlands \\
loes.kruger@ru.nl}
\and
\IEEEauthorblockN{Paul Kobialka\qquad}
\IEEEauthorblockA{
\textit{University of Oslo\;\qquad}\\
Oslo, Norway\qquad \\
paulkob@ifi.uio.no\qquad}
\and
\IEEEauthorblockN{Andrea Pferscher}
\IEEEauthorblockA{
\textit{University of Oslo}\\
Oslo, Norway \\
andreapf@ifi.uio.no}
\linebreakand
\qquad\IEEEauthorblockN{Einar Broch Johnsen}
\IEEEauthorblockA{
\qquad\textit{University of Oslo}\\
\qquad Oslo, Norway\\
\qquad einarj@ifi.uio.no}
\and
\IEEEauthorblockN{Sebastian Junges}
\IEEEauthorblockA{
\textit{Radboud University}\\
Nijmegen, The Netherlands \\
sebastian.junges@ru.nl}
\and
\IEEEauthorblockN{Jurriaan Rot}
\IEEEauthorblockA{
\textit{Radboud University}\\
Nijmegen, The Netherlands \\
jurriaan.rot@ru.nl}
}

\maketitle

\begin{abstract}
  Network protocol fingerprinting is used to identify a protocol implementation by analyzing its input-output behavior. Traditionally, fingerprinting operates under a closed-world assumption, where
  models of all implementations are assumed to be available. However, this assumption is unrealistic in practice. 
  When this assumption does not hold, fingerprinting results in numerous misclassifications without indicating that a model for an implementation is missing.
  Therefore, we introduce an
  open- world variant of the fingerprinting problem, where not all models
  are known in advance. We propose an incremental fingerprinting approach to solve the problem by combining active automata
  learning with closed-world fingerprinting. 
  Our approach quickly determines
  whether the implementation under consideration matches an available model using fingerprinting
  and conformance checking. If no match is found, it learns a new model
  by exploiting the structure of available models.
  We prove the correctness of our approach and improvements in asymptotic complexity compared to naive baselines.
  Moreover, experimental results on a variety of protocols demonstrate a significant
  reduction in misclassifications and interactions
  with these black-boxes.
\end{abstract}

\section{Introduction}
\emph{Fingerprinting} is the problem of identifying a system. 
In this paper, we study how to identify a (network) protocol that is running on a black-box system. 
That is, we want to know the version of the protocol that is used on a device.  
We identify the protocol by the input-output behavior observed while interacting with the device, thus we consider \emph{behavioral fingerprinting}.
Concrete instances of the behavioral fingerprinting problem are to identify the Bluetooth chip used in car keys or the SSH version used in a doorbell camera~\cite{NCCGroupDoorbell,NCCGroup}. 
Determining the protocol version allows us to conclude whether the system is subject to (known) security vulnerabilities.

\myparagraph{The need to fingerprint.}
Bluetooth communication is ubiquitous in modern internet-of-things (IoT) systems. 
In 2024 alone, 4.9 billion
Bluetooth devices were shipped.\footnote{\url{https://www.bluetooth.com/2025-market-update/}}
Given its security-critical role, for example, in the locking system of cars~\cite{NCCGroup}, the Bluetooth protocol demands rigorous implementation. 
This urgency was reinforced by recent vulnerabilities that allowed attackers to forcibly pair earbuds with Google Fast Pair without user consent~\cite{whisperpair2026} or enabled unpaired attackers in proximity to hijack headphones that internally used an Airoha chip~\cite{Airoha}.
In the latter, many headphone and earbud vendors build on top of these vulnerable chips and many of them were unaware that these chips were used in their devices. Fingerprinting methods allow for determining whether a given Bluetooth device is susceptible to such attacks.

\myparagraph{Fingerprinting Finite State Machines.}
In our work, we consider protocols that can be represented by Finite State Machines (FSMs). FSMs can represent various security-relevant network protocols, including TLS~\cite{DBLP:conf/uss/Fiterau-Brostean20,rasoamananaSystematicAutomaticUse2022}, SSH~\cite{Fiterau-Brostean17SSH} and Bluetooth Low Energy (BLE)~\cite{pferscherFingerprintingAnalysisBluetooth2022}. 
The standard assumption for behavioral finterprinting is to assume access to all protocol versions~\cite{DBLP:conf/uss/Fiterau-Brostean20,Fiterau-Brostean17SSH,PferscherA21BLE,rasoamananaSystematicAutomaticUse2022}. 
This assumption is called a \emph{closed-world assumption} and is overly optimistic: There is no exhaustive list of BLE chips or SSH implementations. 
In this paper, we propose studying fingerprinting without a closed-world assumption, that is, \emph{we study fingerprinting in an open world.} Section~\ref{sec:motivation} demonstrates that wrongly assuming a closed world leads to a significant number of misclassifications.

\myparagraph{Fingerprinting in a closed world.}
In a closed world, we are given a \emph{reference set} of (known) FSM \emph{models} and a set of black-box devices (from here onwards: \emph{implementations}). This is a variation on behavioral fingerprinting of a single device and is also known as \emph{group matching}~\cite{shuFormalMethodologyNetwork2011}.
Behavioral fingerprinting techniques similar to \cite{DBLP:conf/ccs/ArgyrosSJKK16} use the reference set to compute a \emph{fingerprint}, a small set of so-called \emph{separating sequences} that together identify any model in the reference set. 
Fingerprinting an implementation in a closed world is easily done by executing the fingerprint on them. 
The result will match any black-box implementation with exactly the matching reference model.  

\myparagraph{Fingerprinting in an open world.}
In an open-world setting, instead, the reference set cannot be assumed to be complete. 
As a result, closed-world fingerprinting techniques may not match with any model (\emph{no match}) or with the wrong model (\emph{a misclassification}). In Section~\ref{sec:motivation}, we show that they tend to misclassify an implementation instead of indicating that there is no matching reference. 
Misclassification due to missing models can be avoided by learning a model of each implementation from scratch using, e.g., Active Automata Learning (AAL), see below. 
However, we demonstrate in Section~\ref{sec:motivation} that this requires an excessive amount of interactions with the black-box protocols. 
This raises a key challenge: 
\begin{mdframed}
How can we limit the number of needed interactions, while keeping the number of misclassifications to a minimum?
\end{mdframed}

\myparagraph{Active Automata Learning.}
To address the above challenge, our approach tightly integrates fingerprinting and AAL to deal with the open-world setting. AAL is a well-established technique for constructing models of black-box systems. 
In fact, if the reference models are not known in advance during closed-world fingerprinting, they are often learned using AAL as a preprocessing step~\cite{DBLP:conf/ccs/ArgyrosSJKK16, janssen2021fingerprinting, PferscherA21BLE, rasoamananaSystematicAutomaticUse2022}. 
These AAL algorithms learn a behavioral model of the so-called System Under Learning (SUL) through interactions\,\cite{steffen2011introduction, Vaandrager17Overview}.
In this work, we consider AAL algorithms designed for deterministic, data-less FSMs equipped with a reliable reset mechanism. 
The state of the art in AAL for such FSMs includes algorithms such as
$L^*$~\cite{angluin1987learning}, TTT~\cite{DBLP:conf/rv/IsbernerHS14} and \lsharp~\cite{VaandragerGRW22}. They construct a hypothesis by targeted interactions with the SUL, where a hypothesis is an FSM. Then they extensively test the hypothesis using conformance checking against the SUL. The conformance check is the main bottleneck as it typically requires millions of interactions for real-world systems. Expert knowledge is often used to terminate the conformance check and the resulting model is assumed correct.\looseness=-1

\myparagraph{Our approach.} 
We present \ifalg, a novel and incremental fingerprinting algorithm to accurately and efficiently identify black-box devices under an open-world assumption. \ifalg matches implementations with any previously learned reference model \emph{and then tests whether this match is not a misclassification}. Only if the implementation cannot be matched, we learn its model. 
Instead of learning a new model from scratch, we use adaptive AAL, which ensures amortized costs by considering similar reference models~\cite{DBLP:conf/tacas/GrocePY02,KrugerJR24StateMatching}. After learning a new model, we add it to the reference set.
Compared to closed-world fingerprinting, incremental fingerprinting is proven to correctly classify all implementations and to only learn a model of an implementation when it is distinct from all previous models given an adequate conformance testing oracle.  
We illustrate its effectiveness on a motivating example and in the empirical evaluation using various common network protocols.\looseness=-1

\myparagraph{Contributions.} In summary, this paper presents:
\begin{enumerate}
\item Formalization of the open-world fingerprinting problem for FSMs and showing the necessity to solve it.
\item The incremental fingerprinting algorithm, called
  \ifalg, that solves open-world fingerprinting problems.
  \item Proofs of the correctness of \ifalg (Thm.\,\ref{thrm:correctnes}) and improved query complexity w.r.t.\ baselines (Thm.\,\ref{thm:comp}).
  \item The efficiency and robustness of \ifalg in exhaustive experiments demonstrated on several network protocols.\looseness=-1
\end{enumerate}

\myparagraph{Outline.}
Section~\ref{sec:motivation} motivates incremental fingerprinting in an open world by example.
The incremental fingerprinting problem is formally introduced in Section~\ref{sec:problem} and the state of the art is presented in Section~\ref{sec:state-of-the-art}. Section~\ref{sec:incremental_fingerprinting} details our novel \ifalg algorithm, which is then evaluated in Section~\ref{sec:experiments}. Sections~\ref{sec:discussion}, \ref{sec:related_work}, and \ref{sec:conclusion} discuss the approach, situate it with respect to related work, and conclude with final insights and future work.


\begin{figure*}[t]
    \begin{subfigure}[b]{0.31\textwidth}
  

\begin{tikzpicture}[->,>=stealth',shorten >=1pt,auto,node distance=1.5cm,main node/.style={circle,draw,font=\sffamily\large\bfseries},
  ]
  \def\xoffset{8mm}
  \node[initial,state,basis] (0) {\treeNodeLabel{$q_0$}};
  \node[state,basis] (1) [below of=0, yshift=-0.5cm] {\treeNodeLabel{$q_1$}};
  \node[state,basis] (2) [right of=1, xshift=0.5cm] {\treeNodeLabel{$q_2$}};
  
  \path[every node/.style={font=\sffamily\scriptsize}]
  (0) edge[loop right] node[right, align=center] {kex/error,\\data/error} (0)
      edge[] node[left] {hello/hello} (1)
  (1) edge[loop left] node[left, align=center] {hello/hello,\\ \textcolor{black}{data/error}} (1)
      edge[bend right] node[below] {kex/kex} (2)
  (2) edge[bend right] node[above, align=center] {hello/hello} (1)
      edge[loop above] node[above, align=center] {data/data,\\kex/kex} (2)
      ;
  \end{tikzpicture}
        \caption{$\M_0$}
    \end{subfigure}
    \begin{subfigure}[b]{0.31\textwidth}
  
      

\begin{tikzpicture}[->,>=stealth',shorten >=1pt,auto,node distance=1.5cm,main node/.style={circle,draw,font=\sffamily\large\bfseries},
  ]
  \def\xoffset{8mm}
  \node[initial,state,basis] (0) {\treeNodeLabel{$r_0$}};
  \node[state,basis] (1) [below of=0, yshift=-0.5cm] {\treeNodeLabel{$r_1$}};
  \node[state,basis] (2) [right of=1, xshift=0.5cm] {\treeNodeLabel{$r_2$}};
  
  \path[every node/.style={font=\sffamily\scriptsize}]
  (0) edge[loop right] node[right, align=center] {kex/error,\\data/error} (0)
      edge[] node[left] {hello/hello} (1)
  (1) edge[loop left] node[left, align=center] {hello/error,\\ \textcolor{black}{data/error}} (1)
      edge[bend right] node[below] {kex/kex} (2)
  (2) edge[bend right] node[above, align=center] {hello/error} (1)
      edge[loop above] node[above, align=center] {data/data,\\kex/error} (2)
      ;
  \end{tikzpicture}
        \caption{$\M_1$}
    \end{subfigure}
    \begin{subfigure}[b]{0.36\textwidth}
        \begin{tikzpicture}[->,>=stealth',shorten >=1pt,auto,node distance=1.5cm,main node/.style={circle,draw,font=\sffamily\large\bfseries},
  ]
  \def\xoffset{8mm}
  \node[initial,state,basis] (0) {\treeNodeLabel{$s_0$}};
  \node[state,basis] (1) [below of=0, yshift=-0.5cm] {\treeNodeLabel{$s_1$}};
  \node[state,basis] (2) [right of=1, xshift=0.5cm] {\treeNodeLabel{$s_2$}};
  
  \path[every node/.style={font=\sffamily\scriptsize}]
  (0) edge[loop right] node[right, align=center] {kex/error,\\data/error} (0)
      edge[] node[left] {hello/hello} (1)
  (1) edge[loop left] node[left, align=center] {hello/error,\\data/error} (1)
      edge[bend right] node[below] {kex/kex} (2)
  (2) edge[] node[right, align=center] {hello/error,\\kex/error} (0)
      edge[loop right] node[right, align=center] {data/data} (2)
      ;
  \end{tikzpicture}
  \vspace{-0.42cm}
        \caption{$\M_2$}
    \end{subfigure}
    \caption{Finite state machines representing simplified TLS protocols. }
    \label{fig:mealies}
\end{figure*}

\section{Overview} \label{sec:motivation}
In this section, we motivate the necessity to account for an open
world in fingerprinting, even when most models are already known. Additionally, we show that
simply learning every model from scratch is prohibitively expensive.
We formalize our models as Finite State Machines (FSMs). We introduce FSMs and separating sequences by
example and refer to Section~\ref{sec:preliminaries} for formal definitions.

\begin{example}[Finite State Machines]
\label{ex:mm}
    Fig.\,\ref{fig:mealies} depicts three FSMs $\M_0$, $\M_1$ and $\M_2$ representing simplified TLS protocols. The models use input alphabet $I = \{\emph{hello}, \emph{kex}, \emph{data}\}$ representing a hello message, key exchange and data sending, respectively. These inputs can generate outputs from the alphabet $O = \{\emph{hello}, \emph{kex}, \emph{data}, \emph{error}\}$. 
    The FSMs behave differently with respect to handling \emph{hello} messages at unexpected times. For example, FSMs $\M_0$ and $\M_1$ are distinct as ${\M_0}$ responds to $ \emph{hello}\cdot\emph{hello}$ with $\emph{hello}\cdot\emph{hello}$, while $\M_1$ responds with $\emph{hello}\cdot\emph{error}$. The input sequence $\emph{hello}\cdot\emph{hello}$ is therefore called a \emph{separating} sequence for ${\M_0}$ and ${\M_1}$.
\end{example}
A common approach to fingerprinting with a set of reference models, represented as FSMs, is to compute a \emph{fingerprint}, i.e., a set of separating sequences that together uniquely identify any of the original FSMs (e.g.,~\cite{rasoamananaSystematicAutomaticUse2022}). \looseness=-1
When such a fingerprint for a set of reference models is executed on an implementation
that is \emph{not} already represented in the set, two outcomes are possible. Ideally, the considered implementation disagrees with each reference model on at least one separating sequence in the fingerprint, thereby indicating that the implementation is new. For example, $\{\emph{hello}\cdot\emph{kex}\cdot\emph{hello}\cdot\emph{hello}\}$ is a fingerprint for the set of models $\{\M_0,\M_1\}$ (Fig.\,\ref{fig:mealies}). If we regard $\M_2$ as an implementation and execute this fingerprint on it, we find
that $\M_2$ is a new implementation.
However, for a different fingerprint, the considered implementation may match a reference model on all fingerprint sequences, even though it represents a distinct system. For instance, if we take instead the fingerprint $\{\emph{hello}\cdot\emph{hello}\}$ for $\{\M_1, \M_2\}$, then $\M_2$ is incorrectly matched with $\M_1$.

\myparagraph{Closed-world fingerprinting in an open world.}
We conduct a motivational experiment to assess how often closed-world fingerprinting methods detect that an implementation behaves differently from all the references.
We use 596 implementations of the TLS protocol with 22 underlying FSMs from~\cite[Section~6.4]{janssen2021fingerprinting} and assume that 21 of the 22 FSMs that underlie the 596 implementations are known. 
Then, we evaluate the performance of closed-world fingerprinting using separating sequences (see Section~\ref{sec:state-of-the-art} for details)
based on the number of misclassifications and implementations that could not be matched to a reference.
We average over $10$ runs, each with one random model removed and shuffling the references before constructing the fingerprint.
We observe that 4.5\% of the implementations are misclassified and all implementations can be matched to some model. 
The number of misclassifications and implementations that cannot be identified grows with the number of missing models. When presented with 11 out of 22 models, 45.9\% of the matches are misclassified and 2.2\% cannot be matched.
These misclassification rates suggest that closed-world fingerprinting methods lack robustness when the references are incomplete. In such cases, they are more likely to misclassify a model rather than recognize the absence of a match.\looseness=-1

\myparagraph{Learning models from scratch.}  
As an alternative approach, the state-of-the-art AAL algorithm~\lsharp~\cite{VaandragerGRW22} can be used to learn a model of each implementation and evaluate the correctness of the learned models. We find that \lsharp requires roughly 2.6 million interactions to learn a model of each black-box protocol implementation. However, this results in an incorrect model for 75.9\% of the implementations.
The high number of incorrectly learned implementations originates from the conformance check being inadequate. Even with a more exhaustive conformance check, where learning the model set requires a total of 14.3 million iterations, this still leads to 32.6\% of the implementations being incorrectly learned.

\myparagraph{\ifalg.}
Algorithmically, we suggest two key adaptations over learning a model of each implementation from scratch. 
First, to avoid relearning previous models, we test whether the implementations match a model that we have previously seen. Second, to enable the reuse of previous models instead of learning from scratch, we consider adaptive AAL~\cite{DBLP:conf/tacas/GrocePY02}. Adaptive learning reuses existing reference models in order to reduce the number of required interactions. These adaptions lead to a two-step approach: First, identify a likely match with the previously seen models. If such a match is found, check conformance between the black-box and this hypothesis model. If no match is found or the match fails the conformance check, learn the model, guided by previously learned models. We exemplify this approach below.

\begin{example}[Incremental Fingerprinting (\ifalg)]
We continue Example~\ref{ex:mm} and consider fingerprinting four black-box implementations, $\Impl_0,~\Impl_1,~\Impl_2$ and $\Impl_3$ such that 
$$\Impl_0 \sim \M_0,\qquad\Impl_1 \sim \M_1 \sim \Impl_3, \qquad\Impl_2 \sim \M_2$$
where $\sim$ denotes behavioral equivalence (see Section~\ref{sec:preliminaries}).
Initially, the set of reference model $\Specs$ is empty, therefore, we learn model $\Spec_0$ representing $\Impl_0$ and add it to $\Specs$.  

Subsequently, we consider $\Impl_1$. We forego fingerprinting, as there is only one model in $\Specs$ and use a conformance check to determine whether $\Spec_0 \sim \Impl_1$. When testing $\emph{hello}\cdot\emph{hello}$, the outputs produced by $\M_0$ and $\M_1$ show that these implementations are distinct. This indicates the need to learn a model of $\Impl_1$, leading to the inclusion of $\Spec_1$ in $\Specs$. 

Next, we consider $\Impl_2$ and derive $\M_0 \nsim \Impl_2$ using fingerprint $\{\emph{hello}\cdot\emph{hello}\}$. Observing that both $\M_1$ and $\Impl_2$ generate the output $\emph{hello}\cdot\emph{error}$, we initially hypothesize that $\M_1 \sim \Impl_2$. However, a conformance check containing sequence $\emph{hello}\cdot\emph{kex}\cdot\emph{hello}\cdot\emph{hello}$ refutes this hypothesis. The learning process is triggered and learned model $\Spec_2$ is added to $\Specs$. 

Finally, when considering $\Impl_3$, we apply the fingerprint $\{\emph{hello}\cdot\emph{kex}\cdot\emph{hello}\cdot\emph{data}\}$, which leads to hypothesis $\Impl_3 \sim \Spec_1$. Because this hypothesis is correct, we terminate after a passed conformance check and conclude $\Impl_3 \sim \Spec_1$.\looseness=-1
\end{example}

Misclassifications are absent in this example but tend to arise in larger models when conformance checks, either after fingerprinting or during learning, fail to cover the complete behavior. When using incremental fingerprinting in the motivational experiments, no misclassifications are produced by the algorithm when starting with 21 reference models. When starting with 11 reference models, incremental fingerprinting beats learning from scratch both in the misclassification rate (0.7\% instead of 32.9\%) and in the number of interactions (3.5 million instead of 14.2 million).


\section{Problem Statement} \label{sec:problem}
In this section, we introduce the preliminaries required to formalize both the fingerprinting problem in a closed world as described in the literature, and the new open world variation. 

\subsection{Preliminaries}
\label{sec:preliminaries}
Throughout this paper, we fix a finite set $I$ of inputs and a finite set $O$ of outputs. 
\begin{definition}
A \emph{Finite State Machine} is a tuple $\M = (Q, q_0, \delta, \lambda)$ with finite set $Q$ of \emph{states}, \emph{initial state} $q_0 \in Q$, \emph{transition function} $\delta\colon Q \times I \to Q$ and \emph{output function} $\lambda\colon Q \times I \to O$.
 \end{definition}
 The transition and output functions are extended to input sequences of length $n \in \mathbb{N}$ as functions $\delta\colon Q \times I^n \to Q$ and $\lambda\colon Q \times I^n \to O^n$. We use superscript $\M$ to refer to elements of an FSM, e.g.,~$Q^{\M}$ and $\delta^{\M}$. 
 We denote the concatenation of inputs and outputs for two sequences $v, v' \in I^*$, respectively $O^*$, as $v \cdot v'$. We write $\lambda(w)$ instead of $\lambda(q_0,w)$ for some $w \in I^*$. We denote the size of an object $S$, such as a set or list, by $|S|$. Given an FSM \M, $|\M|$ refers to $|Q^\M|$. 

\begin{definition} \label{def:eq}
    Given a language $L \subseteq I^*$ and FSMs $\M_0$ and $\M_1$, two states $p \in Q^{\M_0}$ and $q \in Q^{\M_1}$ are $L$\emph{-equivalent}, written as $p \sim_L q$, if $\lambda^{\M_0}(p,w)=\lambda^{\M_1}(q,w)$ for all $w \in L$. 
\end{definition} 
FSMs $\M_0$ and $\M_1$ are $L$-equivalent, written  $\M_0 \sim_L \M_1$, if $q_0^{\M_0}$ and $q_0^{\M_1}$ are $L$-equivalent. The states $p$ and $q$ are \emph{equivalent}, written $p \sim q$, if they are $I^*$-equivalent. Analogously, $\M_0$ and $\M_1$ are equivalent if $q_0^{\M_0} \sim q_0^{\M_1}$. 
Intuitively, separating sequences witness the inequality of FSMs:
\begin{definition}
    Given FSMs $\M_0$ and $\M_1$, a sequence $\sigma \in I^*$ is a \emph{separating sequence} for $\M_0$ and $\M_1$ if $\lambda^{\M_0}(\sigma) \neq \lambda^{\M_1}(\sigma)$. 
    A set of sequences $L \subseteq I^*$ is a \emph{fingerprint} for a set of models \Specs if every pair of non-equivalent models $\M_0, \M_1 \in \Specs$ has a separating sequence in $L$, and every $\sigma \in L$ separates some pair of models $\M_0, \M_1 \in \Specs$.
    \end{definition}

\subsection{Formal Problem Statement}
We define the \emph{fingerprinting problem under a closed-world assumption}, also known as \emph{group matching fingerprinting}~\cite{ShuL11PEFSM}. 

\begin{problemstatement}{Closed-World Fingerprinting}
Given a set of inequivalent models $\Specs$ and a list
of black-box implementations $\Impls \subseteq \Specs$, compute a map $\mu \colon \Impls \to \Specs$ s.t.\ for all $\Impl \in \Impls,~\Spec \in \Specs$:~$\mu(\Impl) = \Spec$ iff $\Impl \sim \Spec$. 
\end{problemstatement}

By the inclusion $\Impls \subseteq \Specs$, we indicate that for every $\Impl \in \Impls$ there exists $\Spec \in \Specs$ such that $\Impl \sim \Spec$.
The assumption $\Impls \subseteq \Specs$ 
ensures that every implementation is equivalent to a reference model.  \label{sec:openproblem}
In Section~\ref{sec:motivation}, we saw that this assumption is often too strict. This motivates the open world variation of fingerprinting, where the set of models is not known a priori. Thereby, building the set \Specs becomes a part of the problem. 

\begin{problemstatement}{Open-World Fingerprinting} \label{ps:owf}
Given a list of black-box implementations \Impls, compute a set of models $\Specs$ and a map $\mu \colon \Impls \to \Specs$ s.t.\ for all $\Impl \in \Impls,~\Spec \in \Specs$:~$\mu(\Impl) = \Spec$ iff $\Impl \sim \Spec$.
\end{problemstatement}

\noindent If an initial set of inequivalent models is available, the problem can easily be adapted to support a ``warm start''.\looseness=-1


\section{State of the Art}
\label{sec:state-of-the-art}
In this section, we discuss the state of the art for fingerprinting, conformance checking and active automata learning. 

\subsection{Fingerprinting Algorithms} 
\label{sec:fingerprint_algs}
Given an implementation, a fingerprinting algorithm returns a potentially matching model from a fixed set of reference models.
An algorithm executes a sequence by running it on the implementation.
Contrary to standard definitions, we assume that the fingerprinting algorithm also returns the set of all executed sequences.

\begin{definition}\label{def:fingerprinting}
    A \emph{fingerprinting algorithm} requires an implementation $\Impl$ and a set of models $\Specs$ as inputs. The algorithm executes a subset $L_F \subseteq I^*$ of the fingerprint for $\Specs$. It returns $L_F$ and $\Spec$ if there is a model $\Spec \in \Specs$ which is the only model that satisfies $\Impl \sim_{L_F} \Spec$; otherwise, it returns $L_F$ and \emph{None}.\looseness=-1
\end{definition}

If the closed-world assumption holds, i.e., for each $\Impl \in \Impls$ there exists $\Spec \in \Specs$ such that $\Impl \sim \Spec$, fingerprinting algorithms always return exactly one model. To solve the fingerprinting problem under a closed-world assumption, we execute a fingerprinting algorithm to each implementation $\Impl \in \Impls$ such that a reference model $\Spec \in \Specs$ is found for which $\Impl \sim \Spec$ holds.\looseness=-1

\myparagraph{Static fingerprinting algorithms} compute a set of separating sequences and run these sequences on implementation \Impl in an arbitrary order, terminating once at most a single model remains~\cite{rasoamananaSystematicAutomaticUse2022, ShuL11PEFSM}. 
Because each separating sequence distinguishes a pair of models in \Specs, and the fingerprint includes one separating sequence for every such pair, executing all sequences on \Impl guarantees the elimination of at least $|\Specs| - 1$ models.\looseness=-1

\myparagraph{Dynamic fingerprinting algorithms} also compute a set of separating sequences but determine on-the-fly which sequence to run next. A simple approach is to only select separating sequences that are guaranteed to rule out at least one of the remaining models. This can be accomplished through, e.g., \emph{adaptive distinguishing graphs} (ADGs)~\cite{janssen2021fingerprinting}. 
An alternative implementation of ADGs might order the separating sequences based on the expected
number of inequivalent models after applying the sequence, following the definition of adaptive distinguishing sequences~\cite{VaandragerGRW22}. We use the latter ADG interpretation throughout this paper. 

\begin{example}
    Suppose we want to construct a fingerprint of references $\M_0, \M_1$ and $\M_2$ from Fig.~\ref{fig:mealies}. The separating sequence $\emph{hello}\cdot\emph{hello}$ distinguishes the pairs $(\M_0, \M_1)$ and $(\M_0, \M_2)$, whereas $\emph{hello}\cdot\emph{kex}\cdot\emph{hello}\cdot\emph{hello}$ separates $(\M_1, \M_2)$. A static fingerprinting algorithm begins by applying the separating sequence for $(\M_0, \M_1)$; if the implementation is not equivalent to $\M_0$, a second sequence is required. The ADG approach described above 
    observes that $\emph{hello}\cdot\emph{kex}\cdot\emph{hello}\cdot\emph{hello}$ distinguishes all models, and thus only requires one sequence during application.
\end{example}

\subsection{Conformance Checking} \label{sec:conformance_algs}
Conformance checking~\cite{DBLP:journals/tse/Chow78, vasilevskii1973failure} studies whether a given model coincides with a black-box implementation. 
Under a closed-world assumption, a fingerprint is sufficient evidence to conclude that the implementation coincides with a model. However, in an open world, the selected model may not actually be equivalent to the implementation. A \emph{conformance query} (CQ) checks if a given implementation $\Impl$ and model $\Spec$ conform, using a conformance checking algorithm.

\begin{definition}\label{def:conformance-checking}
A conformance checking algorithm requires an implementation $\Impl$ and a model $\Spec$ as inputs. The algorithm returns $L_{CQ} \subseteq I^*$ along with a Boolean outcome: \emph{true} if $\Impl \sim_{L_{CQ}} \Spec$ and \emph{false} otherwise.
\end{definition}

We consider three categories of conformance checking: (a) algorithms with strong guarantees, (b) lightweight algorithms, and (c) algorithms providing a trade-off between them. 

\myparagraph{Wp} (in cat.\ a) is an established conformance checking algorithm~\cite{DBLP:journals/tse/FujiwaraBKAG91}. It creates a test suite of input sequences by: (1) accessing all states in the model, (2) performing $k$ input steps from each state, and
    (3) checking whether the expected state has been reached using a separating sequence.
Wp guarantees that if the implementation has at most $k$ more states than the model, then the test suite contains a separating sequence if the implementation and model are inequivalent.
Wp is expensive as it is exponential in $k$.  

\myparagraph{RandomWord} (in cat.\ b) generates random input sequences of a specified length and compares the outputs from the model and implementation~\cite{angluin1987learning}. While random sequences are very cheap to generate, only statistical guarantees can be given. 

\myparagraph{RandomWp} (in cat.\ c) combines Wp and RandomWord based on ideas described in~\cite{DBLP:conf/icfem/SmeenkMVJ15}. It visits all states in the model, performs a \textit{random walk}, and then checks whether the expected state has been reached. It does not give the same guarantees as Wp, but is very effective in finding separating sequences with few queries~\cite{AichernigTW20Benchmarking, GarhewalD23}.

\subsection{Learning Algorithms} \label{sec:learning_algs}
\emph{Active automata learning} (AAL) aims to learn models of black-box systems by systematically providing inputs and observing outputs. AAL algorithms are a natural candidate to learn implementations and incrementally build the reference set in our setting. We refer the reader to the surveys by Howar and Steffen~\cite{HowarS16Overview} and Vaandrager~\cite{Vaandrager17Overview} for an overview of AAL.

\begin{definition}\label{def:aal}
    An AAL algorithm requires an implementation $\Impl$ as input and returns a model $\Spec$ such that $\Impl \sim \Spec$.
\end{definition}

AAL algorithms are often implemented within Angluin's \emph{Minimally Adequate Teacher} (MAT) framework~\cite{angluin1987learning}. In this framework, the learning algorithm has access to a teacher who has perfect knowledge of the \emph{System Under Learning} (SUL). 
The teacher can answer two types of queries: \emph{Output Queries} (OQs) and \emph{Equivalence Queries} (EQs). When asked an OQ with a given input sequence $w$, the teacher returns the output sequence as produced by SUL $\M$, i.e., $\lambda^{\M}(q_0,w)$. When asked an EQ, the teacher answers whether a provided hypothesis model $\mathcal{H}$ is equivalent to SUL $\M$, i.e., whether $\mathcal{H} \sim \M$ holds. If the provided hypothesis is incorrect, the teacher returns a counterexample that witnesses the behavioral difference. The counterexample can be used by the learner to refine the hypothesis.\looseness=-1

\begin{example}
    Suppose we want to learn a model of $\M_0$ from Fig.~\ref{fig:mealies}. A learning algorithm might start by posing OQs: \emph{hello}, \emph{kex} and \emph{data}.
    Based on the teacher's responses, we construct initial hypothesis $\mathcal{H}_0$ shown in Fig.~\ref{fig:hyp}.
    Next, we pose an EQ with $\mathcal{H}_0$. The teacher might respond with counterexample $\emph{hello}\cdot\emph{kex}$, which produces $\emph{hello}\cdot\emph{error}$ in $\mathcal{H}_0$ but $\emph{hello}\cdot\emph{kex}$ in $\M_0$. This discrepancy reveals the existence of a second state. 
    
    To construct the next hypothesis, we need to identify transitions that lead to different states. This can be done using a separating sequence, such as \emph{kex}.\footnote{This example is consistent with the $\lsharp$ algorithm which identifies states by distinguishing them from all but one state, in contrast to algorithms like $L^*$ that aim to establish equivalence.} For instance, to find the target of the transition $\emph{hello}\cdot\emph{kex}$, we pose OQ $\emph{hello}\cdot\emph{kex}\cdot\emph{kex}$. If this path reaches the initial state, then \emph{kex} should yield \emph{error}; if it reaches the second state, it should return \emph{kex}.
    Using \emph{kex} as separating sequence to identify all transitions, we build refined hypothesis $\mathcal{H}_1$. This hypothesis can be refuted by counterexample $\emph{hello}\cdot\emph{kex}\cdot\emph{data}$, which discovers the last state. After identifying each transition once more, we arrive at the correct hypothesis which is equivalent to $\M_0$. Posing an EQ at this point confirms its correctness, and the learning process terminates.\looseness=-1
\end{example}

\begin{figure}[t]
    \centering
    \scalebox{0.95}{\begin{tikzpicture}[->,>=stealth',shorten >=1pt,auto,node distance=1.5cm,main node/.style={circle,draw,font=\sffamily\large\bfseries},
  ]
  \def\xoffset{8mm}
  \node[initial,state,basis] (0) {\treeNodeLabel{$h_0$}};
  
  \path[every node/.style={font=\sffamily\scriptsize}]
  (0) edge[loop above] node[above, align=center] {kex/error,\\data/error,\\hello/hello} (0)
      ;
  \end{tikzpicture}}
    \hspace{1cm}
    \scalebox{0.95}{\begin{tikzpicture}[->,>=stealth',shorten >=1pt,auto,node distance=1.5cm,main node/.style={circle,draw,font=\sffamily\large\bfseries},
    ]
    \def\xoffset{8mm}
    \node[initial,state,basis] (0) {\treeNodeLabel{$h_0$}};
    \node[state,basis] (1) [right of=0, xshift=1cm] {\treeNodeLabel{$h_1$}};;
    
    \path[every node/.style={font=\sffamily\scriptsize}]
    (0) edge[loop above] node[above, align=center] {kex/error,\\data/error} (0)
        edge[] node[above] {hello/hello} (1)
    (1) edge[loop above] node[above, align=center] {hello/hello,\\ data/error,\\kex/kex} (1)
        ;
    \end{tikzpicture}}
    \caption{Hypotheses $\mathcal{H}_0$ and $\mathcal{H}_1$ for SUL $\M_0$. }
    \label{fig:hyp}
\end{figure}

We consider the state-of-the-art algorithm \lsharp~\cite{VaandragerGRW22}. This algorithm efficiently learns a model from scratch by using an efficient data structure to store all interactions with the SUL. \lsharp outperforms the automata learning algorithm $L^*$~\cite{angluin1987learning} and is competitive with algorithms like TTT~\cite{DBLP:conf/rv/IsbernerHS14}.
Additionally, we consider \emph{adaptive} active automata learning~\cite{DBLP:conf/tacas/GrocePY02,DBLP:conf/splc/DamascenoMS19}, in particular the algorithm \adaptivelsharp~\cite{KrugerJR24StateMatching} built on top of \lsharp. Adaptive learning algorithms reuse information from a set of reference models $\Specs$. If these models closely resemble the SUL, the learning process can speed up significantly.
For open-world fingerprinting, the models learned so far can be used as reference models.

\myparagraph{Approximating the EQ.}
Assuming a teacher with perfect knowledge of the SUL, \lsharp and \adaptivelsharp learn the correct model using a number of queries polynomial in the number of states and inputs of the SUL, as well as the length of the longest counterexample. 
However, such a perfect teacher often does not exist as the SUL is a black-box implementation.
Thus, a teacher is often implemented by executing OQs directly on the SUL and approximating the EQs using conformance checking, see Section~\ref{sec:conformance_algs}. When using conformance checking, the learning algorithm returns a model $\Spec$ that is equivalent with the SUL w.r.t.\ all input sequences posed during learning and conformance checking. 

\begin{definition}\label{def:aal-cc}
    An adaptive AAL algorithm using conformance checking requires an implementation $\Impl$, a set of models $\Specs$, and $L_F \subseteq I^*$ as inputs. The algorithm returns a model $\Spec$ and $L_L \subseteq I^*$ such that $\Impl \sim_{L_F \cup L_L} \Spec$.
\end{definition}

In this setting, we allow initialization of the data structure using a set of sequences $L_F$. Data structure initialization using logs is frequently used in AAL to speed up the learning process, as described in~\cite{DBLP:conf/wcre/YangASLHCS19}.


\begin{figure*}[t]
    \begin{adjustbox}{max width=0.99\textwidth}
    \begin{tikzpicture}[node distance=2cm]
    
        \node (full_background) [full_background] {};
        \node (background) [background] {};
        \node (init_spec) [startstop, left of=background, xshift=-
        8.2cm, yshift=1.5cm] {Models $\Specs_0$};
        \node (impl) [startstop, below of=init_spec,yshift=0.5cm] {Implementations $\Impls$};
        \node (specprime) [io, right of=init_spec, xshift=7.5cm] {Models $\Specs$};
        \node (fp) [process, right of=impl, xshift=1.5cm,yshift=-1cm] {Fingerprinting};
        \node (ct) [process, right of=fp, xshift=3cm] {Conformance Query};
        \node (learn) [process, right of=ct, xshift=3cm] {Learning};
        \node[text width=2cm] at (-6.9, 2.3) {$\textsc{IdentifyOrLearn}_\mathcal{C}(\Impl, \Specs)$};
        \node[text width=1cm] at (-11.8, 2.8) {$\textsc{IncrementalFingerprinting}_\mathcal{C}(\Impls, \Specs_0)$};
        \node (out)[right=3.2cm of specprime] {};
    
        \draw [arrow] (init_spec) -- (specprime);
        \draw [arrow] (-6.7,1) -- node[anchor=east] {\Specs} (fp);
        \draw [arrow] (3,1) -- node[anchor=east] {\Specs} (3,-0.5);
        \draw [arrow] (impl) |- node[anchor=north, align=center,xshift=0.5cm] {Implementation \Impl}(fp);
        \draw [arrow] (fp) -- node[anchor=south, align=center] {\Spec, $L_F$} (ct);
        \draw [arrow] (fp) |- ++(0,-0.9cm) -| node[below, xshift=-5cm] {No Model, $L_F$} (learn);
        \draw [arrow] (ct) -- node[anchor=south] {No Model, $L_{CQ}$} (learn);
        \draw [arrow] (ct) -- node[anchor=west, align=left] {\Spec,\\ $\mu(\Impl) \coloneqq \Spec$,\\ $\gamma(\Impl)\coloneqq L_F \cup L_{CQ}$} (-1.65,1);
        \draw [arrow] (3.35,-0.5) -- node[anchor=west, align=left] {$\Specs \coloneqq \Specs \cup \{\Spec\}$,\\ $\mu(\Impl)\coloneqq \Spec$,\\ $\gamma(\Impl) \coloneqq L_F \cup L_{CQ} \cup L_L$} (3.35,1);
        \draw [arrow] (specprime) -- node[anchor=south, align=center, xshift=-0.5cm] {$\Specs$, $\mu$, $\gamma$}(out);
    \end{tikzpicture}
    \end{adjustbox}
    \caption{Overview of Incremental Fingerprinting.}
    \label{fig:overview}
    \end{figure*}

\section{Incremental Fingerprinting}
\label{sec:incremental_fingerprinting}
In this section, we introduce a framework for incremental fingerprinting, combining closed-world fingerprinting and automata learning to solve the open-world fingerprinting problem efficiently and accurately. 
We detail the algorithm, its correctness and complexity. 
We conclude this section with a recommended configuration of the algorithm, which we refer to as \emph{the} incremental fingerprinting algorithm \ifalg. \camerareadyversion{The proofs of all theorems and the complete benchmark results can be
found in the appendix of the extended version of this paper~\cite{Kruger2026OpenWorldFingerprintingArxiv}.}{}

\subsection{Incremental Fingerprinting Algorithm}
Algorithm~\ref{alg:main_alg} lists \textsc{IncrementalFingerprinting} which solves the open-world fingerprinting problem, i.e., it returns $\Specs$ such that for all $\Impl \in \Impls$ there is an equivalent model in $\Spec$.
The algorithm iteratively builds a model set \Specs, and functions $\mu$ and $\gamma$. 
For each $\Impl \in \Impls$, $\mu$ stores an equivalent model $\Spec \in \Specs$, while $\gamma$ records the input sequences $L \subset I^*$ used to identify $\Impl$.
Upon termination, $\mu(\Impl) = \Spec \leftrightarrow \Impl \sim_{\gamma(\Impl)} \Spec$ holds for each implementation $\Impl \in \Impls$.
The workflow of \textsc{IncrementalFingerprinting} is outlined in Fig.~\ref{fig:overview} and the algorithm is listed in Alg.~\ref{alg:main_alg}.

In Lines~1--2 of Alg.~\ref{alg:main_alg}, we initialize $\Specs$ and iterate over all implementations $\Impls$.
In Line~3, we run the algorithm \textsc{IdentifyOrLearn}, discussed below, with the current implementation $\Impl$ and set of references $\Specs$.
The \textsc{IdentifyOrLearn} algorithm either matches $\Impl$ with a reference in \Specs or learns a new model. 
In Lines~4--7, the model $\Spec$ and language $L$ returned by \textsc{IdentifyOrLearn} are used to update $\Specs$, $\mu$ and $\gamma$. When all $\Impl \in \Impls$ are considered, we return $\Specs$, $\gamma$ and $\mu$.

Before discussing \textsc{IdentifyOrLearn}, we explain the behavior of the \textsc{Learn} algorithm used in Lines 1 and 7. $\textsc{Learn}$ requires an implementation $\Impl$, a set of models $\Specs$ and a language $L_F \subseteq I^*$ to initialize the learning data structure. 
Internally, it makes use of several CQs to approximate the EQ. Algorithm $\textsc{Learn}$ returns a model $\Spec$ and language $L_L$ of OQs used during learning, for which $\Impl \sim_{L_L \cup L_F} \Spec$ holds.  

\textsc{IdentifyOrLearn}, presented in Alg.~\ref{alg:identify_or_learn}, either identifies a matching model \Spec in the set of references \Specs, or learns a new model. 
In Line~1, we handle the special case where $\Specs$ is empty and call \textsc{Learn} with $\Specs = \emptyset$ and $L_F = \emptyset$. We return the result of \textsc{Learn}: found model $\Spec$ and language $L_L$.

In Line~2, we use fingerprinting to identify the reference in \Specs that is a candidate for equivalence with implementation \Impl. $\textsc{Fingerprinting}$ requires an implementation $\Impl$ and a set of inequivalent models $\Specs$ as inputs. The algorithm returns the set of executed sequences $L_F$ to avoid reposing the sequences during the CQ or when learning a new model. Additionally, a candidate reference $\Spec \in \Specs$ if $\Spec$ is the only reference in \Specs for which $\Impl \sim_{L_F} \Spec$ holds and \textit{None} otherwise.

In Lines 3--6, we test the candidate reference $\Spec$ using a CQ if fingerprinting resulted in a non-\textit{None} model $\Spec$.
$\textsc{ConfQuery}$ requires an implementation $\Impl$ and a model $\Spec$ as inputs. $\textsc{ConfQuery}$ returns $L_{CQ}$ and a Boolean $b$ set to \textit{true} if $\Impl \sim_{L_{CQ}} \Spec$ and \textit{false} otherwise. After execution, we update $L_F$ to include the input sequences $L_{CQ}$ posed during the CQ.
If the conformance check passes, i.e., $b =$ \textit{true}, we return the found model $\Spec$ and the language $L_F$.
Lines 7--8 handle the case where $\Impl \nsim \Spec$ for all $\Spec \in \Specs$ after fingerprinting or conformance checking. The algorithm $\textsc{Learn}$ in Line~7 receives the set of models $\Specs$ and the language $L_F \cup L_{CQ}$ as inputs. We return the found model $\Spec$ and the language $L_F \cup L_L$.\looseness=-1

Alg.~\ref{alg:main_alg} and \ref{alg:identify_or_learn} are parameterized by a tuple $\mathcal{C}$ which contains the algorithms for fingerprinting ($\textsc{Fingerprinting}_\mathcal{C}$), fingerprinting CQ ($\textsc{ConfQuery}_\mathcal{C}$), learning ($\textsc{Learn}_\mathcal{C}$), and learning CQ ($\textsc{LearningConfQuery}_\mathcal{C}$).

\begin{algorithm}[t]
    \caption{$\textsc{IncrementalFingerprinting}_\mathcal{C}$}\label{alg:main_alg}
    \begin{algorithmic}[1]
    \Require implementations $\Impls$, initial references $\Specs_0$
    \State Initialize $\Specs \gets \Specs_0$
    \For{$\Impl \in \Impls$}
        \State $\Spec, L = \textsc{IdentifyOrLearn}_{\mathcal{C}}(\Impl, \Specs)$
        \State $\Specs \leftarrow \Specs \cup \{\Spec\}$
        \State $\gamma(\Impl) = L$ \Comment{For Thm.~\ref{thrm:correctnes}}
        \State $\mu(\Impl) = \Spec$
    \EndFor{}
    \State \Return \Specs, $\gamma$, $\mu$
    \end{algorithmic}
    \end{algorithm}

\begin{algorithm}[t]
\caption{$\textsc{IdentifyOrLearn}_{\mathcal{C}}$}\label{alg:identify_or_learn}
\begin{algorithmic}[1]
\algnewcommand{\IIf}[1]{\State\algorithmicif\ #1\ \algorithmicthen}
\algnewcommand{\EndIIf}{\unskip}
\Require implementation $\Impl$, references $\Specs$
    \IIf{$\Specs = \emptyset$}
        \Return $\textsc{Learn}_\mathcal{C}(\Impl, \Specs, \emptyset)$ 
    \EndIIf 
    \State $\Spec, L_F \gets \textsc{Fingerprinting}_\mathcal{C}(\Impl, \Specs)$
        \Comment{Section~\ref{sec:fingerprint_algs}}
    \If{$\Spec$ is not \textit{None}} 
        \State $b, L_{CQ} \gets \textsc{ConfQuery}_\mathcal{C}(\Impl, \Spec)$
        \Comment{Section~\ref{sec:conformance_algs}}
        \State $L_F \gets L_F \cup L_{CQ}$
        \IIf{$b$} 
             \Return \Spec, $L_F$
        \EndIIf
    \EndIf
\State $\Spec, L_L \gets \textsc{Learn}_\mathcal{C}(\Impl, \Specs, L_F)$ 
\Comment{Section~\ref{sec:learning_algs}}
\State \Return \Spec, $L_F \cup L_L$
\end{algorithmic}
\end{algorithm}

\subsection{Correctness of Incremental Fingerprinting}
Incremental fingerprinting is correct if $\Impl \sim_{\gamma(\Impl)} \mu(\Impl) \to \Impl \sim \mu(\Impl)$ holds for any implementation \Impl, thus, if we can correctly generalize from the constructed language $\gamma(\Impl)$.
We prove the correctness of \textsc{IncrementalFingerprinting} w.r.t.\ $\Impl \sim_L \Spec$
by formalizing the contracts of \textsc{IdentifyOrLearn} and \textsc{IncrementalFingerprinting}. 
Additionally, we prove that \textsc{IncrementalFingerprinting} is correct when \textsc{ConfQuery} has complete knowledge.

\begin{lemma}
\textsc{IdentifyOrLearn}$_{\mathcal{C}}$~(Alg.~\ref{alg:identify_or_learn}) requires an implementation $\Impl$ and a set of inequivalent models $\Specs$ as inputs. After execution, a model $\Spec$ and a language $L \subseteq I^*$ are returned such that $\Impl \sim_{L} \Spec$ and there is at most one $\Spec' \in \Specs$ for which $\Impl \sim_{L} \Spec'$. Additionally, if \textsc{ConfQuery} and \textsc{LearningConfQuery} in $\mathcal{C}$ are perfect teachers, then $\Impl \sim \Spec$.\looseness=-1
\end{lemma}

\noindent With a perfect teacher, \textsc{IncrementalFingerprinting} precisely solves the open-world fingerprinting problem.

\begin{theorem} \label{thrm:correctnes}
\textsc{IncrementalFingerprinting}$_{\mathcal{C}}$
~(Alg.~\ref{alg:main_alg}) requires a list of implementations $\Impls$ and a set of distinct models $\Specs_0$ as inputs. The algorithm returns $\Specs$, $\gamma$ and $\mu$ such that for $\Impl \in \Impls$ there exists a $\Spec \in \Specs$ for which $\Impl \sim_{\gamma(\Impl)} \Spec$ iff $\mu(\Impl) = \Spec$. 
Additionally, if \textsc{ConfQuery} and \textsc{LearningConfQuery} in $\mathcal{C}$ are perfect teachers, then $\Impl \sim \Spec$.\looseness=-1
\end{theorem}

\subsection{Complexity of Incremental Fingerprinting}
The complexity of learning algorithms is often measured in number of queries used to learn the SUL~\cite{HowarS16Overview,Vaandrager17Overview}. 
For $\lsharp$, learning a model $\M$ has asymptotic query complexity $\mathcal{O}(kn^2 + n \log l)$ where $n = \lvert Q^{\M} \rvert$, $k = \lvert I \rvert$ and $l$ is the length of the longest counterexample~\cite{VaandragerGRW22}. Additionally, we can prove that at most $n$ EQs are required. 
We now analyze the complexity of learning a set of models $\Specs$ representing a list of implementations $\Impls$ when a perfect teacher is available. We first consider repeated application of $\lsharp$ ($\rlsharp$) as a baseline and \textsc{IncrementalFingerprinting} using $\lsharp$ as \textsc{Learn} algorithm.\looseness=-1

\begin{theorem} \label{thm:comp}
    Let $m=\lvert \Specs \rvert$, $i=\lvert \Impls \rvert$. Assume that a perfect teacher is available and that all $\M \in \Specs$ have at most $n$ states, $k$ inputs and counterexamples of length at most $l$. $\rlsharp$ learns the the correct set of models $\Specs$ within $\mathcal{O}(i(kn^2 + n \log l))$ OQs and at most $in$ EQs. \textsc{IncrementalFingerprinting} with $SepSeq$ and $\lsharp$ learns the correct set of models $\Specs$ within $\mathcal{O}(m(kn^2 + n \log l) + im^2)$ OQs and at most $mn + i$ EQs.\looseness=-1
\end{theorem}

Clearly, if $i$ is significantly larger than $m$, our approach asymptotically outperforms $\rlsharp$ when considering the learning queries. \camerareadyversion{}{An overview of the asymptotic complexities is presented in Table~\ref{tab:compl} and proven in App.~\ref{sec:appendix}.}

\begin{table*}[t]
    \caption{Query complexity under a perfect teacher with $m=\lvert \Specs \rvert$, $i=\lvert \Impls \rvert$, $n\leq \lvert \Spec \rvert$ for all $\Spec \in \Specs$, $k=\lvert I \rvert$ and counterexamples of length at most $l$. We assume $i > m$ when fingerprinting black-box implementations of the same protocol.}
    \begin{center}
    \begin{tabular}{|l|ll|}
        \hline
        Algorithm & Maximum OQs & Maximum EQs \\ \hline 
        Repeated $\lsharp$ & $i(kn^2 + n \log l)$ & $in$ \\
        \textsc{IncrementalFingerprinting} with $\lsharp$ & $m(kn^2 + n \log l) + im^2$ & $mn + i$ \\
        Repeated $\adaptivelsharp$ & $i(kn^2 + kmn^2 + n^3m^2) + mn \log l$ & $mn + i-m$ \\
        \textsc{IncrementalFingerprinting} with $\adaptivelsharp$ & $m(kn^2 + kmn^2 + n^3m^2 + n \log l) + im^2$ & $mn + i$\\ \hline
    \end{tabular}
    \label{tab:compl}
\end{center}
\end{table*}

\subsection{Misclassifications}
CQs may yield misclassifications, i.e., they can wrongly conclude $\Impl$ is equivalent to some $\Spec$. As CQs are used within \textsc{IncrementalFingerprinting}, our approach may also yield misclassifications. A misclassification for \textsc{IncrementalFingerprinting} occurs when a model $\mu(\Impl)=\Spec$ while $\Impl \nsim \Spec$, caused by either the CQ on Line~4 of Alg.~\ref{alg:identify_or_learn} or the final learning CQ on Line~1 and 8. However, if incremental fingerprinting starts with a complete set of references, as in the closed-world scenario, no misclassifications can occur.\looseness=-1

\begin{theorem} \label{thm:misc}
    Let $\Impls$ be a list of implementations and $\Specs_0$ a set of inequivalent models such that $\Impls \subseteq \Specs_0$. Executing \textsc{IncrementalFingerprinting}$_{\mathcal{C}}$ with initial references $\Specs_0$ and implementations $\Impls$ returns $\Specs$ and $\mu$ such that $\Specs_0 = \Specs$ and for $\Impl \in \Impls$, $\mu(\Impl) = \Spec$ iff $\Impl \sim \Spec$ for some $\Spec \in \Specs$.
\end{theorem}

This implies that once the algorithm has learned a set of models such that all implementations can be correctly identified, i.e., the reference set is complete, no further misclassifications occur.
The same statement holds for repeated $\adaptivelsharp$ but cannot be proven for repeated $\lsharp$ because no information from previously learned models is used by $\lsharp$. 

\subsection{\ifalg Algorithm}
The incremental fingerprinting algorithm can be instantiated with different algorithms for fingerprinting, conformance checking, learning and conformance checking during learning, as indicated by parameter $\mathcal{C}$. This also exhibits the algorithm's flexibility; components are easily replaced when improved algorithms are designed.
We use \ifalg to refer to the incremental fingerprinting algorithm instantiated with \emph{ADG} for fingerprinting, \emph{RandomWp} to implement both conformance checks, and \adaptivelsharp for learning, as these are all state of the art algorithms.
The effects of different possible instantiations will be investigated in the experimental evaluation of RQ2 in Section~\ref{sec:experiments}.\looseness=-1

\section{Experimental Evaluation} \label{sec:experiments}
In this section, we empirically evaluate the performance of our incremental fingerprinting algorithm \ifalg (Section~\ref{sec:incremental_fingerprinting}), which solves the open-world fingerprinting problem (Section~\ref{sec:problem}). The source code and benchmarks are available online~\cite{OpenWorldFingerprintingZenodo}.\footnote{
    \url{https://github.com/lkruger27/IncrementalFingerprintingOpenWorld}}
In our experiments, we focus on fingerprinting network protocols.
We aim to answer the following research questions:

\begin{description}
    \item[RQ1:] How does \ifalg compare against baselines?
    \item[RQ2:] What algorithmic design choices are crucial for the performance of \ifalg?
    \item[RQ3:] To what extent do misclassifications produced by learning CQs influence misclassifications generated by the fingerprinting CQ? 
\end{description}

\subsection{Benchmarks} \label{sec:benchmarks}
We consider several benchmarks representing network protocol implementations\camerareadyversion{}{, see App.~\ref{sec:model_info} for details.}

\myparagraph{TLS.} Transport Layer Security (TLS) is a well-known security protocol.
We use the 596 implementations and 22 underlying models of mbedTLS and OpenSSL learned and fingerprinted by~\cite{janssen2021fingerprinting}. These models range between 6 and 14 states. 

\myparagraph{SSH.} The Secure Shell Protocol (SSH) is a prominent security protocol of which three implementations have been learned~\cite{Fiterau-Brostean17SSH} using AAL. These models have between 17 and 66 states. We create 17 additional models by applying mutations~\cite{KrugerJR24StateMatching}, such as diverting transitions, removing states, adding states, and changing transition outputs. We consider 100 implementations, where each of the 20 models occurs 5 times.

\myparagraph{BLE.} Bluetooth Low Energy (BLE) is a low-power variant of the Bluetooth communication protocol. BLE devices have previously been learned and fingerprinted~\cite{PferscherA21BLE}. For our evaluation, we use 4 of their models (CC2650, CC2652R1, CYW43455, nRF52832). We extend the BLE model suite by a model of BLE car access systems of a Tesla Model 3~\cite{thesis/Pferscher23} and, additionally, we use models of 3 devices: (1) a proof-of-concept version of CYBLE-416045-02, (2) an updated version of CC2652R1, and (3) Apollo3 Blue. These last three models were learned for this paper\camerareadyversion{}{; learning details can be found in App.~\ref{sec:model_info}}. The 8 models range between 2 and 16 states and use the same 7 inputs. We copy each model 5 times, leading to 40 implementations. \looseness=-1

\myparagraph{BLEDiff.} Furthermore, we consider 6 BLE models learned using the black-box protocol noncompliance checking framework BLEDiff and provided by the authors of~\cite{karim2023blediff}. We make the model input-complete by adding self-loops with a unique output $\epsilon$ for undefined transitions. The 6 models range between 5 and 8 states and have 32 inputs. Due to a difference in the input names and number of inputs, we do not merge BLE and BLEDiff. We copy each model 5 times, resulting in 30 implementations.

\myparagraph{MQTT.} Message Queuing Telemetry Transport (MQTT) is a publish/subscribe protocol often used in IoT~\cite{standard2019mqtt}.
MQTT has been learned and fuzz tested~\cite{AichernigMP21MQTT}. 
We consider an extended set of MQTT brokers containing: HiveMQ 1.3.5, emqx 5.8.6, ejabberd 25.3.0, VerneMQ 2.0.1, Eclipse Mosquitto 2.0.11,
and mochi 2.7.9 with a broadened alphabet.
The models range between 7 and 53 states.
Many of the available MQTT clients include broker logic to improve communication and implement broker behavior already on the client side, e.g., responses to invalid requests.
Therefore, to communicate with the broker, we utilize a custom Java client to test the broker behavior in isolation.
We use the \emph{Py4J} 
library to connect the incremental fingerprinting setup, which is written in Python, to the custom Java client.
Implementations are learned using \lsharp with a state prefix oracle with $10$ walks per state and a walk length of $12$. The state prefix oracle is similar to RandomWp but does not append a separating sequence after the random walk. We copy each model 5 times to model $30$ implementations.

\begin{figure*}[t]
    \begin{subfigure}[b]{0.32\textwidth}
        \centering
        \includegraphics[width=\linewidth]{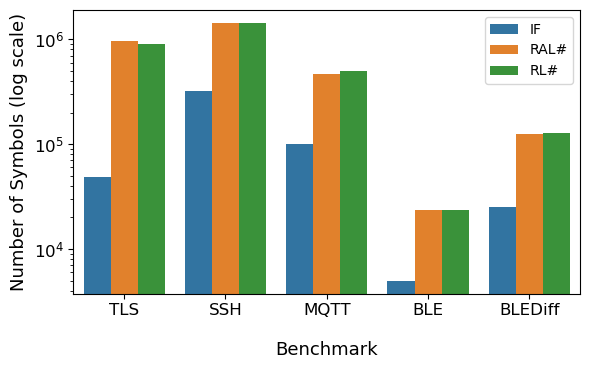}
        \caption{RQ1A: Perfect teacher.}
        \label{fig:1a}
    \end{subfigure}%
    ~ 
    \begin{subfigure}[b]{0.32\textwidth}
        \centering
        \includegraphics[width=\linewidth]{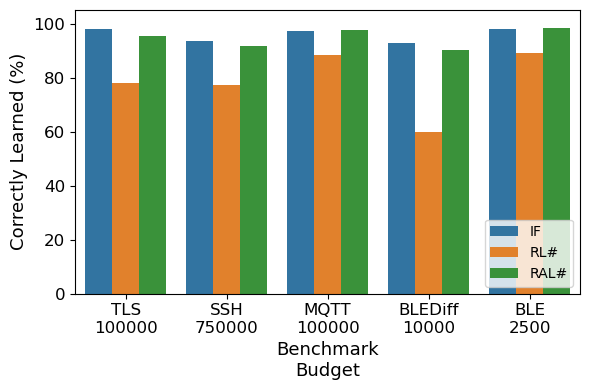}
        \caption{RQ1B: Maximum symbol budget.}
        \label{fig:1b}
    \end{subfigure}
    ~ 
    \begin{subfigure}[b]{0.32\textwidth}
        \centering
        \includegraphics[width=0.99\linewidth]{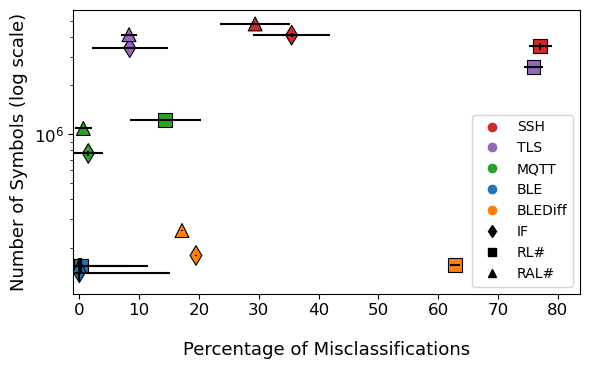}
        \caption{RQ1C: Testing until all tests pass.}
        \label{fig:1c}
    \end{subfigure}
    \caption{Comparison of \ifalg, \rlsharp and \radaptivelsharp for Experiment~\hyperref[exp1]{1}. The colors of bars indicate the algorithms for Figs.~\ref{fig:1a} and \ref{fig:1b}. For Fig.~\ref{fig:1c}, the colors indicate the benchmarks, and the markers indicate the algorithms. The results are averaged over all seeds in all plots. For Fig.~\ref{fig:1c}, the black lines indicate the standard deviations.}
    \label{fig:rq1}
\end{figure*}

\subsection{Experiment Set-up}
We implement \textsc{IncrementalFingerprinting}
using the automata learning library AALpy~\cite{aalpyextended}, using its \adaptivelsharp and \lsharp algorithms. When learning a specific implementation, we rely on a cache containing all OQ responses for fingerprinting, conformance checking, and learning.

\myparagraph{Measurements.}
The performance of the algorithms is measured based on the \emph{number of symbols} (sum of the number of inputs and number of resets) sent to the SUL, as is standard in active automata learning~\cite{VaandragerGRW22}.
Interacting with a black-box system usually requires more time than computing the next OQ, indicating that the number of interactions accurately represents performance. For example, setting up the connection and waiting for network packets to arrive takes a considerable amount of time when interacting with a BLE device.
Additionally, we measure the \emph{percentage of misclassifications} while learning a list of implementations, defined as the ratio of misclassifications to implementations.
To determine whether a misclassification occurs, we check bisimilarity between the ground truth model for an implementation $\Impl$ and $\mu(\Impl)$.
Unlike conventional automata learning experiments, we do not abort the learn process or CQ based on any side information. 
In some experiments, we set a maximum symbol budget. 
The symbol budget is the maximal number of symbols that may be used during learning and testing. In this case,
the learning process can stop either when all tests pass or when the budget runs out. If the budget is exceeded during learning, we return the previous hypothesis. If it happens during testing, we return the hypothesis currently under evaluation. In this scenario, a model is learned correctly if the returned hypothesis is equivalent to the implementation.
We run all experiments with at least 20 different seeds.

\myparagraph{Configuration $\mathcal{C}$.}
We call the naive separating sequence fingerprinting approach from Section~\ref{sec:fingerprint_algs} \textit{SepSeq} and use \textit{ADG} to refer to the second discussed ADG implementation. Inside \ifalg, we use either \adaptivelsharp or \lsharp as discussed in Section~\ref{sec:learning_algs}. We use the following specific settings in our experiments concerning the CQ implementations, see Section~\ref{sec:conformance_algs}:
\begin{itemize}
    \item Wp: The test suite of the Wp method grows exponentially with its parameter $k$. We use $k=2$ in line with conclusions from~\cite{DBLP:conf/icgi/KrugerGV23}
    and as also done in~\cite{GarhewalD23}.
    \item RandomWord: We execute 1000 sequences with lengths ranging from 10 to 30.
    \item RandomWp: The number of walks per state is 100 unless indicated otherwise. Each walk has a random length between 1 and 5 sampled from a uniform distribution.
    \item Budget RandomWp: The random walk length is set to a geometric distribution with minimum length 3 and expected length 8, see~\cite{DBLP:conf/icfem/SmeenkMVJ15}.
\end{itemize}

\myparagraph{Baselines.}
We assume that no models are available at the start of an experiment, which rules out comparisons with closed-world fingerprinting. 
The automata learning algorithm $L^*$ is often used to learn a set of models that is later used in closed-world fingerprinting~\cite{rasoamananaSystematicAutomaticUse2022, pferscherFingerprintingAnalysisBluetooth2022} 
We use \rlsharp (repeated \lsharp) and \radaptivelsharp (repeated \adaptivelsharp) as baselines instead of repeated $RL^*$ as \lsharp outperforms $L^*$~\cite{VaandragerGRW22}. 
In $\rlsharp$, we maintain a set $\Specs$ and a mapping from implementations to models. For every $\Impl$, we run $\lsharp$, add the model if it is not equivalent to any in $\Specs$ and update the mapping. $\radaptivelsharp$ follows the same procedure, using the current $\Specs$ as the reference set.
We note that \radaptivelsharp can be seen as a simplification of \ifalg that omits the fingerprint and subsequent CQ.\looseness=-1

\subsection{Experiment 1: \ifalg vs Baselines \radaptivelsharp and \rlsharp}  \label{exp1}
To answer RQ1, we compare the performance of \ifalg
to baselines \radaptivelsharp and \rlsharp. Given the teacher's significant impact on the performance of \ifalg and the baselines, we examine RQ1 across three scenarios. To this end, we address the following subquestions: How does \ifalg compare against baselines when using 
\begin{enumerate}
    \item A perfect teacher,
    \item RandomWp and a maximum symbol budget, and
    \item RandomWp until all tests pass?
\end{enumerate}

\myparagraph{Results.}
The results for Experiment 1 are depicted in Fig.~\ref{fig:rq1}, per subquestion. Fig.~\ref{fig:1a} shows the number of symbols (log-scaled) sent to the SUL while learning the set of models using a perfect teacher; lower is better. 
Fig.~\ref{fig:1b} shows the percentage of correctly learned models
when learning is terminated after exceeding the budget which is included on the x-axis; higher is better. 
Finally, Fig.~\ref{fig:1c} shows both the number of interactions (log-scaled) and the number of misclassifications. The interactions in Fig.~\ref{fig:1c} are not limited by a budget, instead the experiment ends when the CQ indicates that all tests passed.
This last plot can be used to determine the trade-off between misclassifications and number of interactions; low and left is best.
We perform an additional experiment for RQ1c varying the number of walks per state in RandomWp for SSH and TLS, displayed in Fig.~\ref{fig:ALvIFw_SSH} and \ref{fig:ALvIFw_TLS} \camerareadyversion{in the Appendix}{(App.~\ref{app:add})}.

\begin{problemstatement}{Answer RQ1}
When a perfect teacher is available, \ifalg performs almost an order of magnitude better than the baselines across all benchmarks. 
When a maximum budget is used, \ifalg and \radaptivelsharp are more accurate than \rlsharp. 
When testing until all tests pass, \ifalg and \radaptivelsharp provide a better trade-off between few interactions and few misclassifications compared to \rlsharp.
\end{problemstatement}

\myparagraph{On \rlsharp.} The baseline \rlsharp is not a reasonable solution to open-world fingerprinting as it leads to an excessive number of interactions with the system under a perfect teacher compared to \ifalg (Fig.~\ref{fig:1a}). 
Moreover, \rlsharp leads to an exceptionally high misclassification rate compared to \ifalg and \radaptivelsharp when using a maximal budget or when testing until convergence (Fig.~\ref{fig:1b}). The high misclassification rate is especially noticeable in Fig.~\ref{fig:1c}, \rlsharp has a misclassification of 75.9\% for TLS while \ifalg and \radaptivelsharp have a misclassification rate under 9\%.

\myparagraph{On \ifalg vs \radaptivelsharp}
When considering a perfect teacher (Fig.~\ref{fig:1a}), \ifalg clearly outperforms \radaptivelsharp. However, when such a teacher is not available (Figs.~\ref{fig:1b} and \ref{fig:1c}), the trade-off between \ifalg and \radaptivelsharp is more nuanced: \ifalg often requires fewer symbols than \radaptivelsharp while \radaptivelsharp often produces fewer misclassifications. 
 
Additionally, it can be observed that the same CQ implementation for all benchmarks leads to widely different misclassification rates in Fig.~\ref{fig:1c}. This shows that to obtain an accurate model, it is essential that the CQ is well-configured. 
We hypothesize that \radaptivelsharp has fewer misclassifications because relearning is more effective at revealing distinct implementations than some CQ configurations.
Therefore, we performed additional tests with a more exhaustive CQ. In Fig.~\ref{fig:ALvIFw_SSH} for SSH, the difference in misclassification for \ifalg and \radaptivelsharp diminishes. Moreover, the same experiment for TLS (Fig.~\ref{fig:ALvIFw_TLS}) reveals that \ifalg outperforms \radaptivelsharp for TLS when the CQ is exhaustive enough. Thus, the trade-off between \ifalg and \radaptivelsharp is benchmark and resource specific.

\subsection{Experiment 2: Ablation Study} \label{exp:2}
To answer RQ2, we perform an ablation study to check whether other configurations for \ifalg are better performing and to gain insights into the effect of the individual algorithms on the overall performance. We compare all combinations of the algorithms mentioned in Section~\ref{sec:state-of-the-art}:
\begin{itemize}[leftmargin=*]
    \item \textsc{Fingerprinting}: \textit{SepSeq} and \textit{ADG},
    \item \textsc{ConfQuery}: \textit{Wp2}, \textit{RandomWp100} and \textit{RandomWord1000},
    \item \textsc{Learn}: \lsharp and \adaptivelsharp.
\end{itemize}

\begin{figure*}[t]
    \centering
    \includegraphics[width=0.99\linewidth]{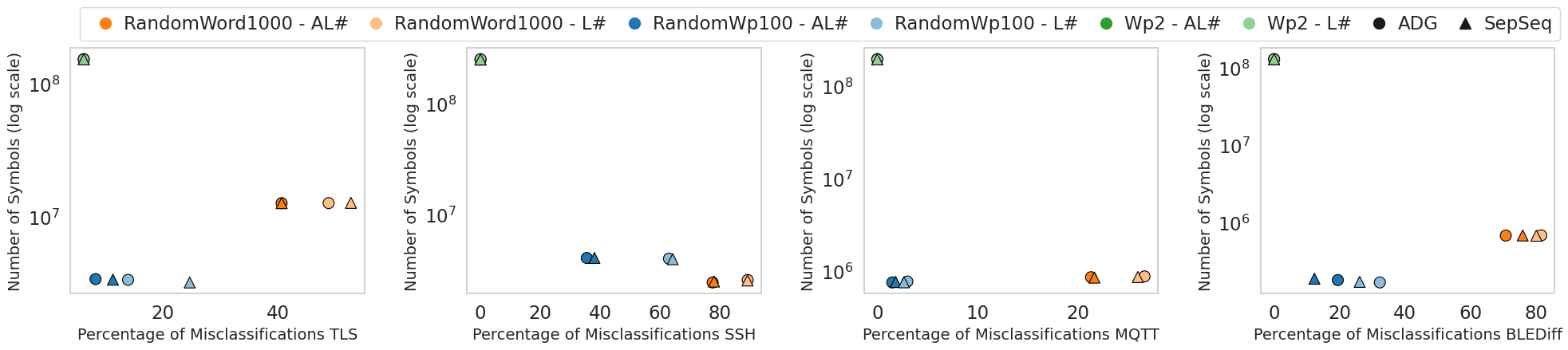}
    \caption{Comparison of different algorithms for the components of incremental fingerprinting for Experiment~\hyperref[exp:2]{2}. In each subplot, color shows CQ, opacity the learning algorithm, and marker the fingerprinting algorithm.}
    \label{fig:rq2}
\end{figure*}

\myparagraph{Results.}
Fig.~\ref{fig:rq2} contains subplots for the TLS, SSH, MQTT, and BLEDiff benchmarks. The results for BLE are presented in Table~\ref{tab:exp2_add} in \camerareadyversion{the Appendix}{App.~\ref{sec:appendix_experimental_evaluation}}, as all BLE models are well differentiable. 
The axis interpretation matches Experiment~\hyperref[exp1]{1c}: lower and left is better. Fig.~\ref{fig:FCQvLCQ} \camerareadyversion{in the Appendix}{in App.~\ref{app:add}} depicts an additional experiment where fingerprinting and learning use different CQ implementations.

\begin{problemstatement}{Answer RQ2} 
The selected CQ algorithm has the biggest impact on the performance, with \textit{RandomWp100} providing the best trade-off between misclassifications and number of symbols. Additionally, we conclude that \textit{ADG} outperforms \textit{SepSeq} and \adaptivelsharp outperforms \lsharp.  These results justify the choices for \ifalg.
\end{problemstatement}

\myparagraph{On ADG vs SepSeq.} 
\textit{ADG} usually leads to fewer misclassifications; the circle marker is more to the left compared to the triangle marker. However, in BLEDiff \textit{SepSeq} outperforms \textit{ADG}. 
The number of interactions is largely dictated by CQ, rendering the fingerprinting algorithm's influence negligible.

\myparagraph{On \adaptivelsharp vs \lsharp: Misclassifications.}
We observe that using \adaptivelsharp during learning leads to fewer misclassifications compared to \lsharp; the opaque color is more to the left. We hypothesize that this is because \adaptivelsharp tests whether states in the known models are also present in the current implementation, leading to bigger and more accurate models. This explanation is in line with Experiment~\hyperref[exp1]{1c}. 

\myparagraph{On \adaptivelsharp vs \lsharp: Interactions.}
The number of interactions does not seem to be impacted as much when using \adaptivelsharp over \lsharp. This can be explained by the fact that learning only occurs when the implementation does not match any of the models in \Specs, which happens 27\% of the time for TLS and 20\% of the time for the other benchmarks. 
When only considering learning symbols, \adaptivelsharp usually requires fewer symbols than \lsharp\camerareadyversion{}{ (Table~\ref{tab:exp2_add}, App.~\ref{sec:appendix_experimental_evaluation})}. For example, incremental fingerprinting with \lsharp, \textit{RandomWp100} and \textit{ADG} requires 25.6\% more learning symbols compared to \adaptivelsharp with the same fingerprinting and CQ settings on average. 

\myparagraph{On the CQ algorithm.}
\textit{RandomWp100} has fewer misclassifications compared to \textit{RandomWord1000} while using the same number of interactions, indicating that \textit{RandomWp100} should be preferred over \textit{RandomWord1000}. \textit{Wp2} produces almost no misclassifications but requires significantly more interactions. The data points for \textit{Wp2} with \ifalg-\adaptivelsharp are hidden behind the data points for \ifalg-\lsharp.
If the number of misclassifications should be minimal, \textit{Wp2} may be preferable over to \textit{RandomWp100}.

\myparagraph{On the number of RandomWp tests.}
In Fig.~\ref{fig:FCQvLCQ} \camerareadyversion{(Appendix)}{(App.~\ref{app:add})}, we experiment with combinations of \textit{RandomWp25}, \textit{RandomWp50} and \textit{RandomWp100} during the fingerprinting CQ (FCQ) and learning CQ (LCQ). As expected, using \textit{RandomWp100} during the FCQ and LCQ produces the fewest misclassifications. Performing a less exhaustive CQ during either the FCQ or LCQ leads to more misclassifications but fewer symbols. The exact trade-off varies depending on the benchmark.

\subsection{Experiment 3: Propagation of Misclassifications} \label{exp:3}
To evaluate how misclassifications produced by learning affect misclassifications produced by fingerprinting (RQ3), we begin by distinguishing two categories of misclassifications, defined by the termination point of Alg.~\ref{alg:identify_or_learn}.
\begin{description}
    \item[FCQ misclassifications:] Terminating with an incorrect $\Spec$ on Line 6, i.e., with the CQ after fingerprinting.
    \item[LCQ misclassifications:] Terminating with an incorrect $\Spec$ on Line 8, i.e., with the CQ within learning.
\end{description}
We consider a configuration of \ifalg with \adaptivelsharp, \textit{ADG} and \textit{RandomWp100} for the FCQ. For the LCQ, we vary between \textit{RandomWp25}, \textit{RandomWp50} and \textit{RandomWp100}. We deliberately use weak CQs to ensure the effect of LCQ misclassifications on FCQ misclassifications is visible.

\begin{figure}[t]
    \centering
    \includegraphics[width=0.99\linewidth]{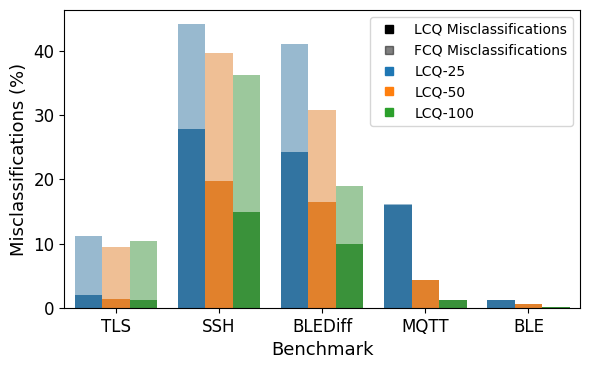}
    \caption{Influence of the LCQ misclassifications on the FCQ misclassifications for Experiment~\hyperref[exp:3]{3}. Colors show the \textsc{LearningConfQuery}. LCQ errors are shown in opaque color and FCQ errors are layered transparently above in the stacked bars.}
    \label{fig:rq3}
\end{figure}

\myparagraph{Results.}
The influence of the LCQ misclassification rate on the FCQ misclassification rate is depicted in Fig.~\ref{fig:rq3}.
Colors indicate the implementation of \textsc{LearningConfQuery}. Misclassification rates are shown as stacked bars: LCQ misclassifications in solid color, FCQ misclassifications transparently.

\begin{problemstatement}{Answer RQ3}
Both CQs are essential for a low misclassification rate. A more exhaustive LCQ usually increases the number of correct models and may reduce the misclassifications during the FCQ.\looseness=-1
\end{problemstatement}

\myparagraph{Discussion.}
The results indicate that a more exhaustive LCQ influences both the LCQ misclassifications and its downstream effects on the FCQ misclassifications.
First, we note that for MQTT and BLE the FCQ is sufficiently exhaustive to detect wrongly learned models, as seen by the non-existent FCQ misclassifications.
For the BLEDiff benchmark, we observe a disproportional increase in correct models as FCQ misclassifications are prevented when fewer incorrect models are provided.
For SSH, we find that a less LCQ misclassification translates to a lower number of overall misclassifications, but increases the number of FCQ misclassifications; both CQs cannot uncover the wrong matches.
In the TLS benchmark, there is no clear trend; the number of FCQ misclassifications fluctuates, while the LCQ misclassification rate marginally improves with more tests.\looseness=-1


\section{Discussion}
\label{sec:discussion}

We discuss limitations of the proposed approach and our experiments, explore a commonly studied problem variation in machine learning, and emphasize the trade-off between misclassifications and performance. 

\myparagraph{Finite state machine learning.}
Our approach requires the ability to learn Finite State Machines (FSMs) representing implementations with active automata learning, which means we assume that interaction with the System Under Learning (SUL) is possible and that the SUL can be represented as an FSM. Such FSMs abstract away from data and timing information.
Many network protocols can be represented as FSMs in theory but exhibit non-determinism in practice due to, e.g., packet loss. It would be interesting to use non-deterministic learning algorithms during incremental fingerprinting. However, the currently available non-deterministic automata learning algorithms are rather pragmatic, inefficient or make assumptions on the type of non-determinism, see e.g. \cite{el2010learning, KhaliliT14, pacharoen2013active}. 
However, when better suited non-deterministic learning algorithms or efficient algorithms for richer types of automata get developed, such as extended FSM which allow (user-entered) data values or FSMs with timers, they can easily be integrated into our incremental fingerprinting framework. 
While tools such as Nmap fingerprinting~\cite{lyon2009nmap} and ssh-audit~\cite{ssh-audit} can identify implementations with only a few probing sequences, they rely on a closed-world assumption and are typically tailored to specific protocol families. In contrast, state-machine learning is agnostic to both the protocol and its implementation.
\looseness=-1

\myparagraph{Black-box learning.}
We do not use side information in the form of source code, documentation, or standards in our approach; we assume a completely black-box scenario. On the one hand, this means we do not rely on side information which allows us to learn models of proprietary software like BLE. On the other hand, many protocol implementations are open-source and, thus, their source code is freely available and could have been used to improve the performance of automata learning and the CQ~\cite{DBLP:conf/forte/ElkindGPQ06}. 
In future work, we plan to investigate gray-box open-world fingerprinting, where, for example, the source code for a subset of the implementations is accessible, or the source code is available but the run-time configuration remains unknown.
\looseness=-1

\myparagraph{Acceptable misclassification percentage.}
Misclassifying a black-box implementations leads to the false belief that it behaves like a certain reference model. Using this flawed reference model to assess vulnerabilities can then lead to false conclusions, such as assuming the implementation is secure when it is not. Thus, preferably an incremental fingerprinting algorithm leads to a 100\% accuracy rate. We aimed for 90\% accuracy when using \ifalg in Experiment~\hyperref[exp1]{1b}, as 100\% typically requires significantly more symbols (see Experiment~\hyperref[exp:2]{2}, 99\% with Wp).

\myparagraph{Duplicate distribution.} 
For all benchmarks except TLS, no realistic distribution of the number of implementations per unique model is available. 
As we focused on using \ifalg for learning a list of implementations with diverse models and duplications, we included five copies of each model in the benchmark sets. \looseness=-1
When repeating Experiment~\hyperref[exp1]{1c} with different numbers of copies per unique model of MQTT, we find that the gain of \ifalg increases with the number of copies, see Table~\ref{tab:different_duplicates} in \camerareadyversion{the Appendix}{App.~\ref{app:add}}. However, this experiment uses a constant number of copies per unique model while in TLS, some of the 22 models have no duplicates while other have 99 duplicates. It would be interesting to gain further insights into the distribution of duplicates and their likeliness in real-world legacy systems and investigate their impact of the performance of \ifalg.

\myparagraph{Implementation clustering.} 
Machine learning approaches for fingerprinting often consider a variation of the fingerprinting problem where equivalent implementations are clustered without building a model set, see e.g.~\cite{marzaniMobileAppFingerprinting2023,msadek2019iot,sabahi2024encrypted,wang2024characterizing}.
We call this variant the \emph{implementation clustering} problem.
A solution to the open-world fingerprinting problem trivially solves the implementation clustering problem.
To compute whether implementations $\Impl$ and $\Impl'$ are equivalent, the returned mapping $\mu$ alone is sufficient as $\mu(\Impl)=\Spec = \mu(\Impl')$ implies $\Impl \sim \Impl'$. 
The reverse does not generally hold.
To build the reference set \Specs from a given set of equivalent implementations, a representative model from each set must be learned.
Further, the set of models returned by open-world fingerprinting favors the explainability of the solution, as all models are individually available and can be analyzed, which is not possible when only implementation clusters are available.

\myparagraph{Passive vs active.}
When active learning is not possible, supervised and unsupervised learning techniques can be used, depending on whether the problem is assumed to be closed or open world.
In a closed world, the training set contains collected pairs of implementations and ground-truth models from which a supervised learner generalizes.
In the open-world variant, there is no such representative training set, and unsupervised learning needs to be used to identify equivalent implementations~\cite{sanchezSurveyDeviceBehavior2021}.
Independent of the setting, the data used to feed those expensive a priori training processes often takes several months or years to collect~\cite{wang2024characterizing,celosia_fingerprinting_2019,marzaniMobileAppFingerprinting2023}.
In contrast, the most costly operation in incremental fingerprinting, the learning of a new model, is only triggered when such a model is actually encountered.

\myparagraph{Incremental fingerprinting in practice.}
Given a list of black-box devices, incremental fingerprinting can be used to build an initial reference set. As new devices appear, they can be identified at a relatively low cost. Many IoT systems integrate BLE devices for which firmware updates can be pulled. In such cases, the incremental approach can be integrated into the IoT system to automatically maintain behavioral models for analysis and identify when potentially harmful devices connect.

\myparagraph{Configuring the CQ.}
We evaluated algorithms on benchmarks with known ground-truth models, allowing misclassification rates to be computed. In practice, ground-truth models are unavailable due to black-box assumptions.
When using incremental fingerprinting instantiated as \ifalg, (\emph{ADG} for fingerprinting, \emph{RandomWp} for CQ, and \adaptivelsharp for learning), it is essential that \emph{RandomWp} is well configured as the performance of \ifalg depend first and foremost on the CQ oracle used.
We propose the following steps to configure the CQ:
Initialize RandomWp with a walk length of 5 and 100 walks per state, based on the results from Experiment~\hyperref[exp1]{1c}. 
Pick one of the black-box implementations to be fingerprinted.
Learn the implementation multiple times over different seeds and double the number of walks per state whenever the learned models vary until all runs stabilize and produce the same model. If behavior is clearly missing from the model, increment the walk length by one.
Repeat this procedure for a few of the other black-box implementations, starting from the configuration that was deemed acceptable for the previous implementation.
The CQ configuration leads to an empirically reasonable trade-off. To get fewer misclassifications, we recommend a more exhaustive CQ such as the Wp-method.

\myparagraph{Exploitation potential.}
State machine learning has proven to be a useful tool for revealing deviations in implementations from the corresponding protocol standards~\cite{DBLP:conf/icst/TapplerAB17,Fiterau-Brostean17SSH,PferscherA21BLE}. 
For example, learned models of Bluetooth implementations could uncover logical errors in the pairing procedure, as was recently exploited in the forced pairing behavior of Google Fast Pair earbuds~\cite{whisperpair2026}, which allowed pairing with any nearby device.
Furthermore, state machine learning has also been proposed for detecting security vulnerabilities~\cite{DBLP:conf/icst/HossenGR11}. This approach is often referred to in the literature as \emph{protocol state fuzzing} and has successfully detected security issues in communication protocols such as TLS~\cite{DBLP:conf/uss/RuiterP15} and DTLS~\cite{DBLP:conf/uss/Fiterau-Brostean20}. 
However, these approaches often rely on the manual analysis of the learned models. In our framework, we could flag models that have known vulnerabilities and check whether an implementation matches any of these models.

We could automate the detection of security vulnerabilities by analyzing learned models using testing or verification techniques. Based on the idea of differential testing for detecting security vulnerabilities~\cite{DBLP:conf/sp/SivakornAPKJ17}, one technique could be to examine differences between the learned models as indicators for possible security vulnerabilities. Such integration can be incorporated directly into the CQ within the \textsc{IdentifyOrLearn} procedure (Alg.~\ref{alg:identify_or_learn}) of our \ifalg framework. For example, we could include input sequences in our test suite that test for the acceptance of invalid hostnames in X.509 certificates. Another approach would be to apply model checking to verify specific properties of the learned models, similar to~\cite{DBLP:conf/cav/Fiterau-Brostean16,Fiterau-Brostean17SSH}. 
These properties could verify whether the models contain a path that bypasses authentication or enables a protocol version downgrade. Note that this extension does not violate the open-world assumption, since test suites and properties only depend on the investigated protocols and are independent of the implementations.
In addition, we can incorporate possible attacks into our learning procedure by applying concepts from learning-based fuzzing~\cite{DBLP:conf/nfm/PferscherA22,AichernigMP21MQTT} to capture the behavior of the implementation under unexpected inputs.

\section{Related Work} \label{sec:related_work}
Fingerprinting is extensively researched for finding known security vulnerabilities in black-box systems.
The survey by Sánchez~\etal summarizes fingerprinting using statistical methods~\cite{sanchezSurveyDeviceBehavior2021}, and Alrabaee~\etal fingerprinting for binary code~\cite{alrabaeeSurveyBinaryCode2023}.
Previous works by Li~\etal and Wang~\cite{DBLP:conf/uss/LiZWL0ZM22,DBLP:conf/sp/Wang20} highlight limitations of closed-world fingerprinting and propose open-world methods for Android apps and websites. These rely on passively captured traffic flows. In contrast, we focus on network protocol fingerprinting via active interactions with a black-box.
In the sequel, we discuss related work on fingerprinting via active automata learning (AAL) and fingerprinting using machine learning.

The most relevant works use AAL to learn a model of the implementations under a closed-world assumption and then perform fingerprinting on the learned models.
Fingerprinting using AAL was initially proposed by Shu~\etal~\cite{ShuL11PEFSM} as an efficient alternative to passive automata learning which requires massive logs~\cite{celosia_fingerprinting_2019}.
In the past, several protocols have been fingerprinted, e.g.~\cite{pferscherFingerprintingAnalysisBluetooth2022,janssen2021fingerprinting,rasoamananaSystematicAutomaticUse2022}.
Pferscher~\etal demonstrates that AAL can be used to learn models of BLE devices, revealing safety-critical behavior, and how these models can be used to fingerprint the learned BLE devices.
Karim~\etal present \emph{BLEDiff}, a black-box compliance checking tool for BLE devices based on deviations between individually learned models \cite{karim2023blediff}.
Janssen demonstrates how TLS stacks can be learned with AAL and then compares different methods for generating fingerprints~\cite{janssen2021fingerprinting}.
Further, Rasoamanana~\etal demonstrates how the TLS stack can be efficiently fingerprinted~\cite{rasoamananaSystematicAutomaticUse2022}.
All of those works use adapted versions of $L^*$, targeted at the specific protocol.
Unlike our work, each new model, even if related to previously learned models, is learned again from scratch.
Our work proposes incremental fingerprinting to solve the fingerprinting problem in an open world.
Incremental fingerprinting only learns a new model if it is determined to be different from all previously seen models, and then performs adaptive learning to make optimal use of the knowledge stored in the previous model.
Fingerprinting of closely related families of models has been studied by
Damasceno~\etal~\cite{damascenoFamilyBasedFingerprintAnalysis2022}, who propose an efficient approach that models whole families as \emph{Featured Finite State Machines} constrained over versions, instead of individual machines. 
By knowing which implementations are related, they assume a closed world.

Recent work utilizes machine learning and stochastic learning to fingerprint black-box implementations. 
Marzani~\etal use passive automata learning and stochastic learning for fingerprinting versions of apps based on their network communication~\cite{marzaniMobileAppFingerprinting2023}.
Further, Wang~\etal fingerprint the platform of a video stream by using machine learning~\cite{wang2024characterizing}, 
and Msadek~\etal fingerprint IoT devices~\cite{msadek2019iot}.
Sabahi-Kaviani~\etal use machine learning to compute the alphabet for automata learning algorithms to classify encrypted traffic~\cite{sabahi2024encrypted}.
All of these approaches can only solve the implementation clustering problem, they can not build a model set \Specs.
In comparison, our work uses incremental fingerprinting with adaptive learning to detect equivalent implementations, but also supports further analysis by learning behavioral models of black-box systems.

\section{Conclusion}
\label{sec:conclusion}
Identifying the network protocol version running on a device allows to assess whether the device is susceptible to known security flaws.
Fingerprinting is often done under a closed-world assumption, which implies that all devices match one of a curated set of known reference models.
In an open world, this set is not complete, and it is not known which models are missing. Learning these models on-the-fly poses challenges in terms of resources and accuracy. We formalized the problem of beavioral open-world fingerprinting and present incremental fingerprinting (\ifalg) to address these challenges by integrating closed-world fingerprinting and active automata learning. The experiments show that \ifalg improves significantly over the state of the art.\looseness=-1

\myparagraph{Future work.} We want to evaluate \ifalg on a more extensive set of network protocol implementations, as well as on models that go beyond software protocols, such as large legacy systems. 
Additionally, we plan to explore gray-box open-world fingerprinting by including side information to improve the learning performance. Side information, such as source code or logs, can help answer output queries without querying the system, or can be used to initialize the learning data structure.
Finally, we will explore combining incremental fingerprinting with machine learning techniques from the literature to conclude equivalence of the fingerprinted version and protocol implementation faster.

\printbibliography
\camerareadyversion{\appendix
\section{Appendix}
\begin{table}[ht]
    \centering
    \caption{Misclassification results for the motivational experiment in Section~\ref{sec:motivation}.}
    \label{tab:motivational_new}
    \begin{adjustbox}{width=0.99\linewidth}
    \begin{tabular}{ll|rrr}
        \toprule
        Algorithm & $|$Initial Models$|$ & Correct Models & Misclassifications & No Matches \\
\midrule
$\rlsharp$: RandomWp100 & 0 &  143.7 - 24.1\% & 452.3 - 75.9\% & 0 - 0.0\% \\
$\rlsharp$: RandomWp500 & 0 &  401.6 - 67.4\% & 194.4 - 32.6\% & 0 - 0.0\% \\
\midrule
Fingerprint: SepSeq & 11 & 309.4 - 51.9\% & 273.6 - 45.9\% & 13.0 - 2.2\% \\
\ifalg & 11 & 591.7 - 99.3\% & 4.3 - 0.7\% & 0.0 - 0.0\% \\
\midrule
Fingerprint: SepSeq & 21 & 568.9 - 95.5\% & 27.1 - 4.5\% & 0.0 - 0.0\% \\
\ifalg & 21 & 595.8 - 100.0\% & 0.2 - 0.0\% & 0.0 - 0.0\% \\
\midrule
Fingerprint: SepSeq & 22 & 596.0 - 100.0\% & 0.0 - 0.0\% & 0.0 - 0.0\% \\
\ifalg & 22 & 596.0 - 100.0\% & 0.0 - 0.0\% & 0.0 - 0.0\% \\
\bottomrule
        \end{tabular}
        \end{adjustbox}
\end{table}

\begin{table}[ht]
    \centering
    \caption{Learning results for the motivational experiment in Section~\ref{sec:motivation}. Conformance symbols cover only the CQ symbols after fingerprinting; learning‑phase symbols belong to the learning symbols.}
    \begin{adjustbox}{width=0.99\linewidth}
    \begin{tabular}{ll|rrrr}
        \toprule
        Algorithm & $|$Initial Models$|$ & Fingerprint Symbols & Conformance Symbols & Learn Symbols & Total Symbols \\
\midrule
$\rlsharp$: RandomWp100 & 0 & 0 & 0 & 2606357 & 2606357\\
$\rlsharp$: RandomWp500 & 0 & 0 & 0 & 14266503 & 14266503\\
\midrule
Fingerprint: SepSeq & 11 & 7275 & 0 & 0 & 7275 \\
\ifalg & 11 & 8415 & 3399028 & 100276 & 3507719 \\
\midrule
Fingerprint: SepSeq & 21 & 10388 & 0 & 0 & 10388 \\
\ifalg & 21 & 8270 & 3485626 & 7210 & 3501105 \\
\midrule
Fingerprint: SepSeq & 22 & 10371 & 0 & 0 & 10371 \\
\ifalg & 22 & 8287 & 3491649 & 0 & 3499936 \\
\bottomrule
        \end{tabular}
        \end{adjustbox}
\end{table}

\begin{table}[ht]
    \centering
    \caption{BLE results for Experiment 2.}
    \label{tab:exp2_add}
    \begin{adjustbox}{width=0.99\linewidth}
\begin{tabular}{ll|r|rrrr}
\toprule
Benchmark & Components & Correct Models & Fingerprint Symbols & CQ Symbols & Learn Symbols & Total Symbols \\
\midrule
BLE & ADG - RandomWord1000 - \adaptivelsharp & 40.0 - 100.0\% & 176 & 672146 & 173516 & 845838 \\
BLE & ADG - RandomWord1000 - \lsharp & 40.0 - 100.0\% & 176 & 672009 & 173161 & 845346 \\
BLE & ADG - RandomWp100 - \adaptivelsharp & 40.0 - 100.0\% & 176 & 109314 & 31154 & 140644 \\
BLE & ADG - RandomWp100 - \lsharp & 40.0 - 100.0\% & 176 & 109300 & 30970 & 140446 \\
BLE & ADG - WpK - \adaptivelsharp & 40.0 - 100.0\% & 176 & 1178870 & 294846 & 1473892 \\
BLE & ADG - WpK - \lsharp & 40.0 - 100.0\% & 176 & 1178870 & 294661 & 1473707 \\
BLE & SepSeq - RandomWord1000 - \adaptivelsharp & 40.0 - 100.0\% & 228 & 672434 & 173226 & 845888 \\
BLE & SepSeq - RandomWord1000 - \lsharp & 40.0 - 100.0\% & 228 & 671884 & 173239 & 845351 \\
BLE & SepSeq - RandomWp100 - \adaptivelsharp & 40.0 - 100.0\% & 228 & 109292 & 31141 & 140661 \\
BLE & SepSeq - RandomWp100 - \lsharp & 40.0 - 100.0\% & 228 & 109252 & 30957 & 140437 \\
BLE & SepSeq - WpK - \adaptivelsharp & 40.0 - 100.0\% & 228 & 1178872 & 294833 & 1473933 \\
BLE & SepSeq - WpK - \lsharp & 40.0 - 100.0\% & 228 & 1178872 & 294568 & 1473668 \\
\bottomrule
\end{tabular}
\end{adjustbox}
\end{table}

\begin{table}[ht]
    \caption{Results for MQTT when using a different number of copies per unique model. All algorithms use RandomWp100 and each experiment is repeated 5 times.}
    \label{tab:different_duplicates}
    \centering
    \begin{adjustbox}{width=0.99\linewidth}
    \begin{tabular}{ll|r|rrrr}
\toprule
Algorithm & Copies & Correct Models & Fingerprint Symbols & CQ Symbols & Learn Symbols & Total Symbols \\
\midrule
\rlsharp & 0 & 4.8 - 80.0\% & 0 & 0 & 245147 & 245147 \\
\radaptivelsharp & 0 & 5.6 - 93.3\% & 0 & 0 & 233369 & 233369 \\
\ifalg & 0 & 5.6 - 93.3\% & 15 & 89 & 238834 & 238938 \\
\midrule
\rlsharp & 2 & 15.8 - 87.8\% & 0 & 0 & 732097 & 732097 \\
\radaptivelsharp & 2 & 17.4 - 96.7\% & 0 & 0 & 658528 & 658528 \\
\ifalg & 2 & 16.8 - 93.3\% & 101 & 222131 & 308863 & 531096 \\
\midrule
\rlsharp & 4 & 26.6 - 88.7\% & 0 & 0 & 1221398 & 1221398 \\
\radaptivelsharp & 4 & 30.0 - 100.0\% & 0 & 0 & 1092100 & 1092100 \\
\ifalg & 4 & 29.2 - 97.3\% & 215 & 493824 & 291786 & 785824 \\
\midrule
\rlsharp & 6 & 36.6 - 87.1\% & 0 & 0 & 1704670 & 1704670 \\
\radaptivelsharp & 6 & 41.6 - 99.0\% & 0 & 0 & 1516457 & 1516457 \\
\ifalg & 6 & 42.0 - 100.0\% & 243 & 768878 & 238334 & 1007455 \\
\midrule
\rlsharp & 8 & 46.4 - 85.9\% & 0 & 0 & 2173474 & 2173474 \\
\radaptivelsharp & 8 & 53.4 - 98.9\% & 0 & 0 & 1943352 & 1943352 \\
\ifalg & 8 & 53.6 - 99.3\% & 358 & 1023326 & 238891 & 1262575 \\
\midrule
\rlsharp & 10 & 58.0 - 87.9\% & 0 & 0 & 2680479 & 2680479 \\
\radaptivelsharp & 10 & 64.6 - 97.9\% & 0 & 0 & 2367425 & 2367425 \\
\ifalg & 10 & 65.4 - 99.1\% & 492 & 1270763 & 256849 & 1528103 \\
\bottomrule
\end{tabular}
\end{adjustbox}
\end{table}

\begin{figure}[ht]
    \centering
    \includegraphics[width=0.9\linewidth]{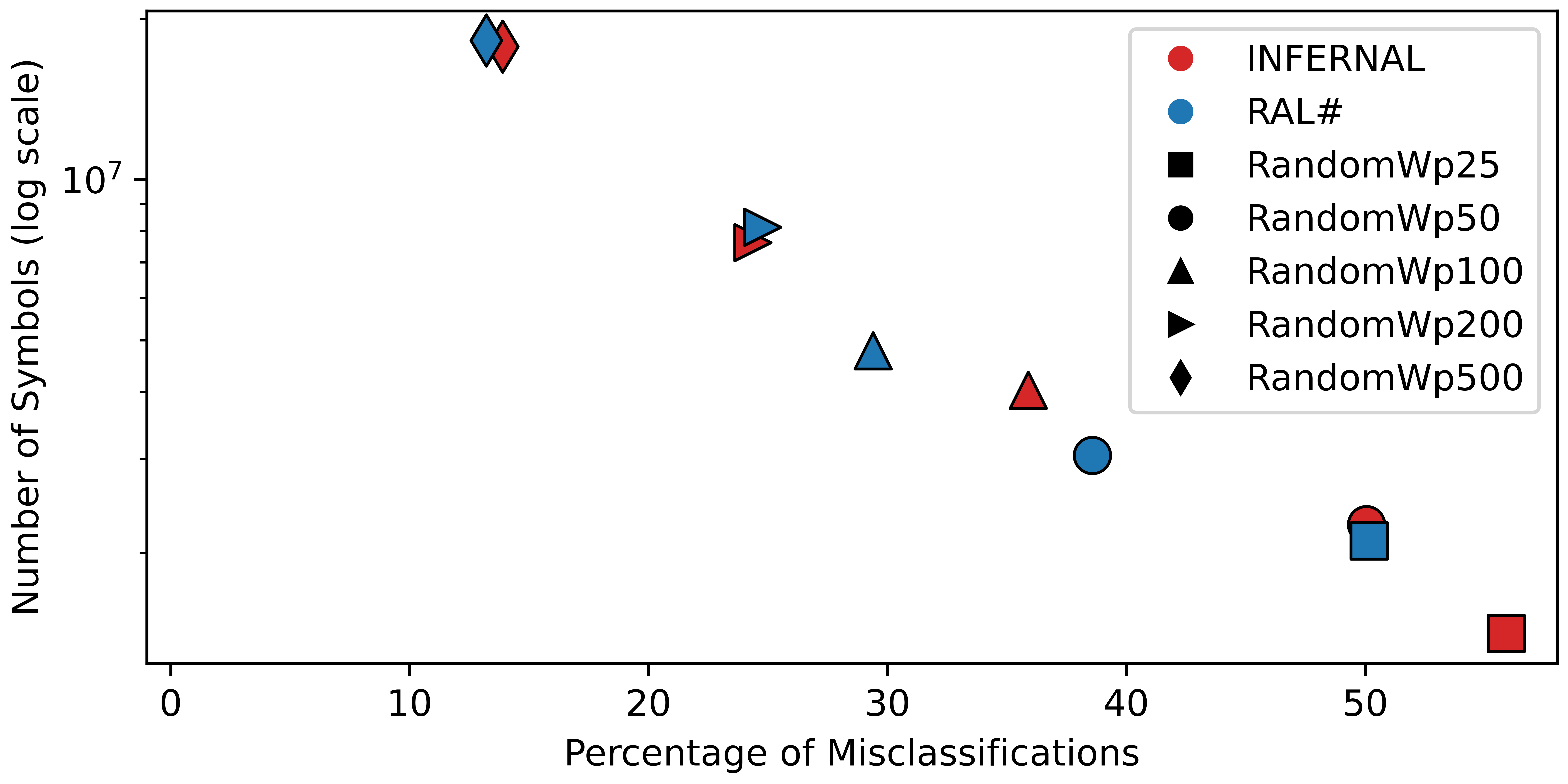}
    \caption{Additional experiment for RQ1. Comparison of \radaptivelsharp and \ifalg for SSH.}
    \label{fig:ALvIFw_SSH}
\end{figure}

\begin{figure}[ht]
    \centering
    \includegraphics[width=0.9\linewidth]{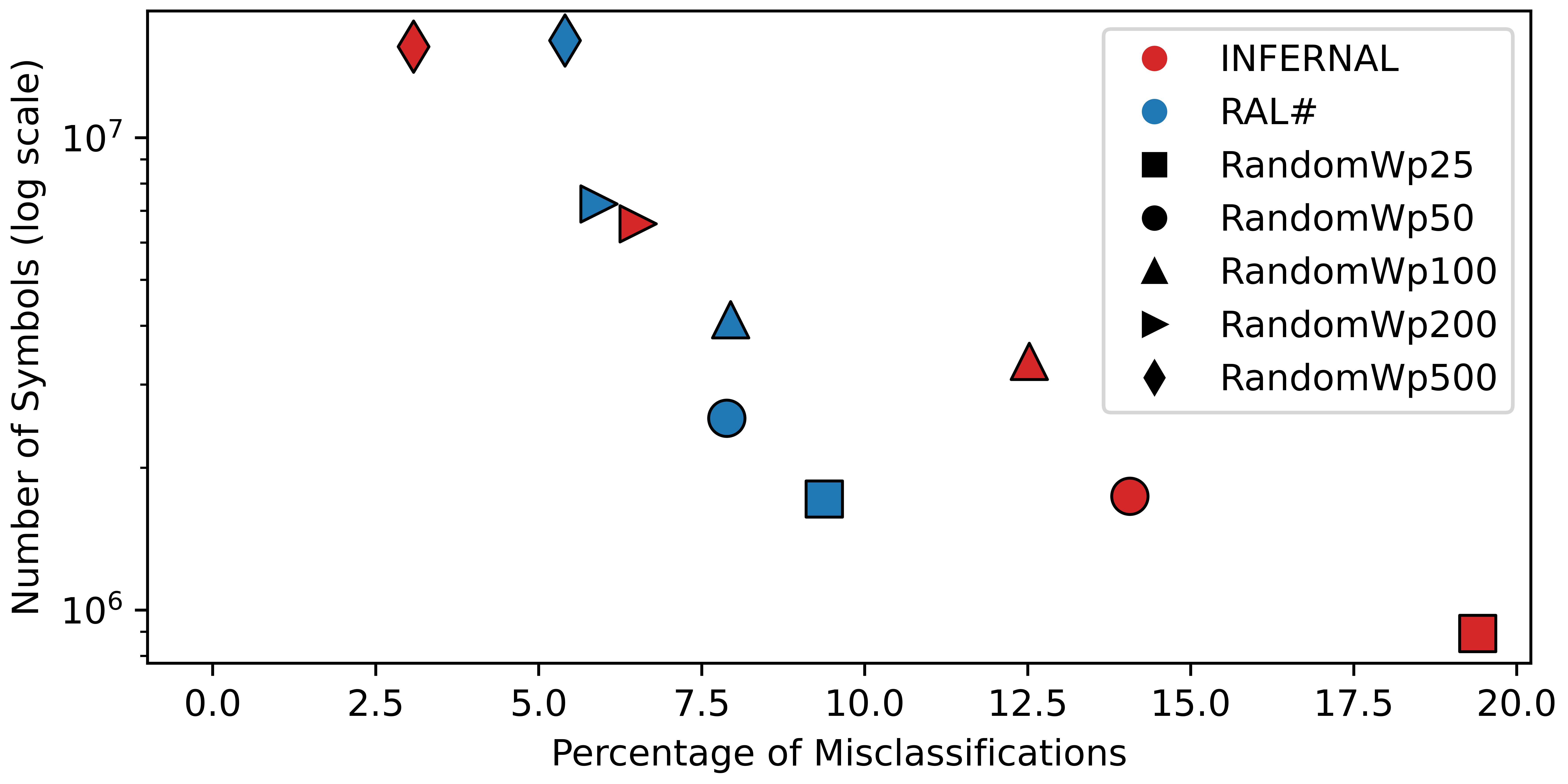}
    \caption{Additional experiment for RQ1. Comparison of \radaptivelsharp and \ifalg for TLS.}
    \label{fig:ALvIFw_TLS}
\end{figure}

\begin{figure}[ht]
    \centering
    \includegraphics[width=0.9\linewidth]{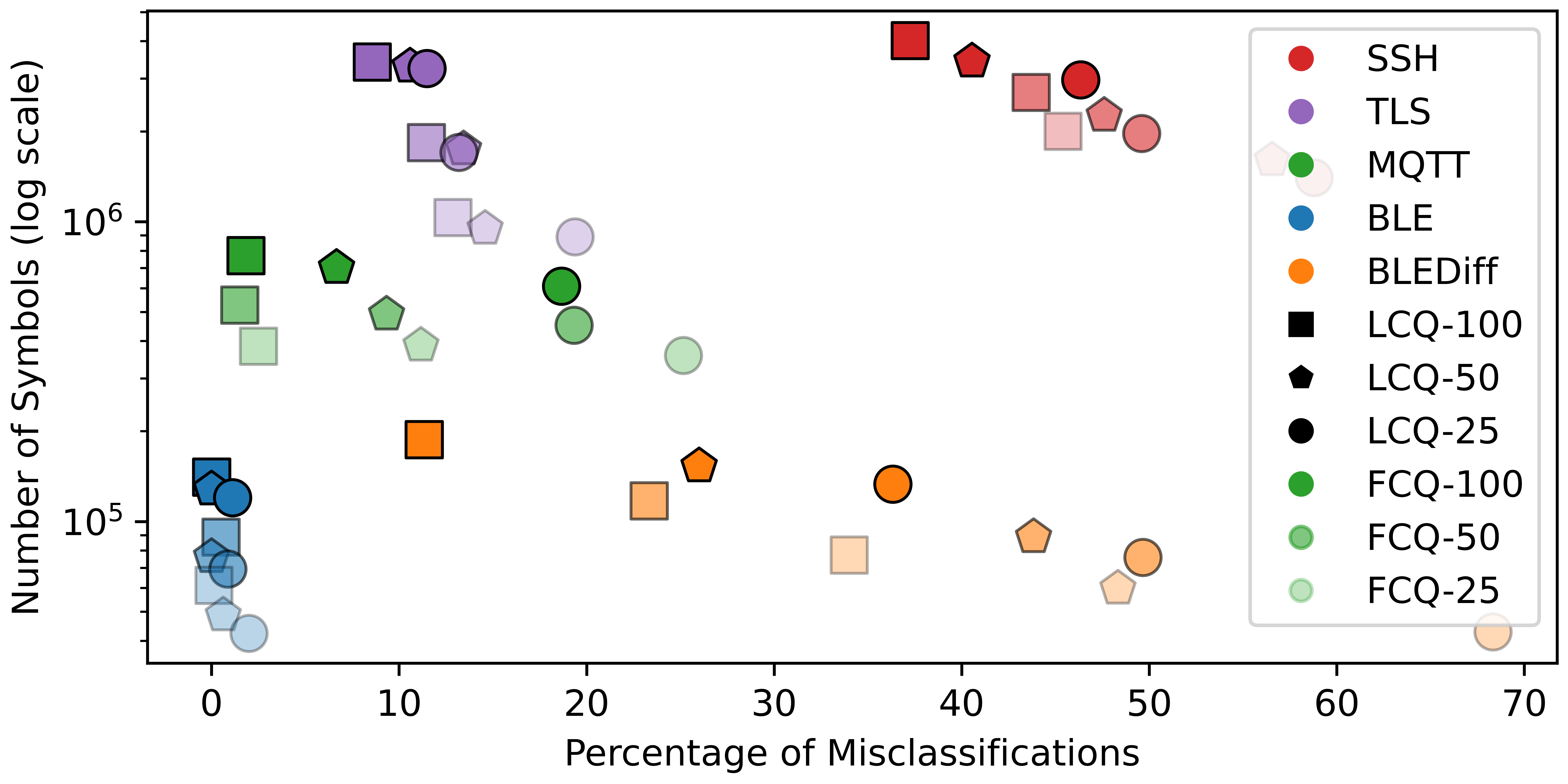}
    \caption{Additional experiment for RQ2. Comparison of the performance of different CQs for fingerprinting and learning.}
    \label{fig:FCQvLCQ}
\end{figure}

}{\newpage

\onecolumn
\appendix

\section{Appendix} \label{sec:appendix}

\subsection{Proof of Theorems}
\textbf{Proof of Theorem 1.} Before diving into the proofs, we reiterate the contracts of the components described in Section~\ref{sec:incremental_fingerprinting}.

\begin{description}
    \item[\textsc{Fingerprinting}] Algorithm \textsc{Fingerprinting} requires an implementation $\Impl$ and a set of models $\Specs$ as inputs. The algorithm executes a subset $L_F \subseteq I^*$ of the fingerprint for $\Specs$. It returns $L_F$ and $\Spec$ if there is a model $\Spec \in \Specs$ which is the only model that satisfies $\Impl \sim_{L_F} \Spec$; otherwise, it returns $L_F$ and \emph{None}.
    \item[\textsc{ConfQuery}] Algorithm \textsc{ConfQuery} requires an implementation $\Impl$ and a model $\Spec$ as inputs. The algorithm returns $L_{CQ} \subseteq I^*$ along with a Boolean outcome: \emph{true} if $\Impl \sim_{L_{CQ}} \Spec$ and \emph{false} otherwise.
    \item[\textsc{Learn}] Algorithm \textsc{Learn} requires an implementation $\Impl$, a set of models $\Specs$ and $L_F \subseteq I^*$ as inputs. The algorithm returns a model $\Spec$ and $L_L \subseteq I^*$ such that $\Impl \sim_{L_F \cup L_L} \Spec$.
\end{description} 
~\\
Under perfect teachers for \textsc{ConfQuery} and \textsc{LearningConfQuery}, we assume the following contracts:
\begin{description}
    \item[\textsc{ConfQuery}] Algorithm \textsc{ConfQuery} requires an implementation $\Impl$ and a model $\Spec$ as inputs. The algorithm returns $L_{CQ} = \emptyset$ and \textit{true} if $\Impl \sim \Spec$ and \textit{false} otherwise.
    \item[\textsc{Learn}] Algorithm \textsc{Learn} requires an implementation $\Impl$ and a set of models $\Specs$ and $L_F \subseteq I^*$ as inputs. The algorithm returns a model $\Spec$ such that $\Impl \sim \Spec$ and $L_L \subseteq I^*$.
\end{description}
Note that the bookkeeping with $L_F$ and $L_L$ is not strictly necessary under a perfect teacher. Moreover, recall that our set of models only includes models that are distinct, i.e., all models are inequivalent.

~\newline
\textbf{Lemma 1.} \textsc{IdentifyOrLearn}$_{\mathcal{C}}$~(Alg.~\ref{alg:identify_or_learn}) requires an implementation $\Impl$ and a set of inequivalent models $\Specs$ as inputs. After execution, a model $\Spec$ and a language $L \subseteq I^*$ are returned such that $\Impl \sim_{L} \Spec$ and there is at most one $\Spec' \in \Specs$ for which $\Impl \sim_{L} \Spec'$. Additionally, if \textsc{ConfQuery} and \textsc{LearningConfQuery} in $\mathcal{C}$ are perfect teachers, then $\Impl \sim \Spec$.
\begin{proof}
    We follow the flow of Alg.~\ref{alg:identify_or_learn} throughout the proof. First, we perform a case distinction based on the size of $|\Specs|$.
    \begin{enumerate}
        \item If $\Specs = \emptyset$, we return \textsc{Learn}($\Impl, \emptyset, \emptyset)$ on Line 1. Following the contract of \textsc{Learn}, we know that a language $L_L \subseteq I^*$ and a model $\Spec$ are returned such that $\Impl \sim_{L_L} \Spec$. Additionally, since $\Specs = \emptyset$ it must hold that $\Spec \notin \Specs$. Moreover, $\Impl \sim_{L} \Spec$ because $L = L_L \cup L_F = L_L \cup \emptyset = L_L$. Thus, there exists a model $\Spec$ with $\Impl \sim_L \Spec$ which is not in $\Specs$. We return $\Spec$ and $L_L$.

        \item If $\Specs$ is not empty, we perform \textsc{Fingerprinting} according to Line 2. \textsc{Fingerprinting} is guaranteed to return at most one $\Spec \in \Specs$ for which $\Impl \sim_{L_F} \Spec$ holds with $L_F \subseteq I^*$. We perform a case distinction based on whether a model is returned or not:
        \begin{enumerate}
            \item One model $\Spec$ is returned and we enter the if-statement starting on Line 3. For this model, we know $\Impl \sim_{L_F} \Spec$ holds. Therefore, we execute \textsc{ConfQuery} and obtain $L_{CQ} \subseteq I^*$ and boolean $b$ with value \textit{true} if $\Impl \sim_{L_{CQ}} \Spec$ and \textit{false} otherwise. We set $L_F$ to $L_F \cup L_{CQ}$ in Line 5 and perform a case distinction based on boolean $b$ in Line 6.
            \begin{enumerate}
                \item $b = \textit{true}$, which indicates $\Impl \sim_{L_{CQ}} \Spec$. When combining this with $\Impl \sim_{L_F} \Spec$, we derive $\Impl \sim_{L_F \cup L_{CQ}} \Spec$. We terminate the algorithm with \Spec, $L = L_F \cup L_{CQ}$ on Line 6. Thus, there exists exactly one $\Spec \in \Specs$ with $\Impl \sim_L \Spec$. 
                
                \item $b = \textit{false}$, which indicates $\Impl \nsim_{L_{CQ}} \Spec$. Because $L_F$ is set to $L_F \cup L_{CQ}$, it holds that $\Impl \nsim_{L_F} \Spec$ for all $\Spec' \in \Specs$. Because we do not enter the if-branch in Line 6, we go to Line 7 and run \textsc{Learn}. \textsc{Learn} returns a model $\Spec$ and $L_L, L_{LCQ} \subseteq I^*$ such that $\Impl \sim_{L_F \cup L_L \cup L_{LCQ}} \Spec$. We terminate the algorithm on Line 8 with $\Spec$ and $L = L_F \cup L_L \cup L_{LCQ}$. Because $\Impl \nsim_{L_F} \Spec'$ for all $\Spec' \in \Specs$ and $\Impl \sim_{L_F \cup L_L \cup L_{LCQ}} \Spec$, it must be the case that $\Spec \notin \Specs$. Thus, there exists a model $\Spec$ with $\Impl \sim_L \Spec$ which is not in $\Specs$. 
            \end{enumerate}
        \item \textit{None} is returned after \textsc{Fingerprinting}, which implies that for all $\Spec \in \Specs$, $\Impl \nsim_{L_F} \Spec$. We go to Line 7 and use the reasoning from 2aii to conclude that there exists a model $\Spec$ with $\Impl \sim_L \Spec$ which is not in $\Specs$. We return $\Spec'$ and $L_F$.
    \end{enumerate}
    \end{enumerate}
    
    \noindent The cases above cover all possible ways to terminate the algorithm \textsc{IdentifyOrLearn}. In all cases, we find a model $\Spec$ such that $\Impl \sim_L \Spec$. When terminating after the conformance check $\Spec \in \Specs$ holds, while terminating after learning guarantees $\Spec \notin \Specs$. Thus, in all cases, there is at most one $\Spec \in \Specs$ for which $\Impl \sim_L \Spec$ holds.

    ~\newline
    Next, we prove: if $\mathcal{C}$ contains perfect teachers \textsc{ConfQuery} and \textsc{LearningConfQuery}, then $\Impl \sim \Spec$.
    \begin{itemize}
        \item From the perfect teacher contracts, we know that \textsc{Learn} using a perfect teacher terminates with $\Impl \sim \Spec$. Combining this information with case 1, 2aii and 2b, it follows that we terminate with $\Impl \sim \Spec'$ for some $\Spec' \notin \Specs$. 
        \item For case 2ai, we remark that a perfect teacher for \textsc{ConfQuery} terminates with $\Impl \sim \Spec$, proving that in this case there exist one $\Spec \in \Specs$ such that $\Impl \sim \Spec$.
    \end{itemize}
    Thus, in both cases $\Impl \sim \Spec$ holds if $\mathcal{C}$ contains perfect teachers for \textsc{ConfQuery} and \textsc{LearningConfQuery}.
\end{proof}

\clearpage
\textbf{Theorem 1.}
\textsc{IncrementalFingerprinting}$_{\mathcal{C}}$
~(Alg.~\ref{alg:main_alg}) requires a list of implementations $\Impls$ and a set of distinct models $\Specs_0$ as inputs. The algorithm returns $\Specs$, $\gamma$ and $\mu$ such that for $\Impl \in \Impls$ there exists a $\Spec \in \Specs$ for which $\Impl \sim_{\gamma(\Impl)} \Spec$ iff $\mu(\Impl) = \Spec$. 
Additionally, if \textsc{ConfQuery} and \textsc{LearningConfQuery} in $\mathcal{C}$ are perfect teachers, then $\Impl \sim \Spec$.
\begin{proof}
    For each $\Impl \in \Impls$, \textsc{IdentifyOrLearn} is guaranteed to return a model $\Spec$ and $L \subseteq I^*$ such that $\Impl \sim_{L} \Spec$ and there is at most one such $\Spec \in \Specs$. 
    %
    Statement $\mu(\Impl)=\Spec \iff \Impl \sim_{\gamma(\Impl)} \Spec$ holds by construction because we update $\Specs$, $\gamma$ and $\mu$ according to the output of \textsc{IdentifyOrLearn} which guarantees $\Impl \sim_{L} \Spec$.

\end{proof}

\textbf{Proof of Theorem 2.} Let $m=\lvert \Specs \rvert$, $i=\lvert \Impls \rvert$. Assume that a perfect teacher is available and that all $\M \in \Specs$ have at most $n$ states, $k$ inputs and counterexamples of length at most $l$. $\rlsharp$ learns the the correct set of models $\Specs$ within $\mathcal{O}(i(kn^2 + n \log l))$ OQs and at most $in$ EQs. \textsc{IncrementalFingerprinting} with $SepSeq$ and $\lsharp$ learns the correct set of models $\Specs$ within $\mathcal{O}(m(kn^2 + n \log l) + im^2)$ OQs and at most $mn + i$ EQs.
\begin{proof}
First, we consider repeated application of $\lsharp$. Because there are $i$ implementations and learning one model using $\lsharp$ requires $\mathcal{O}(kn^2 + n \log l)$ OQs and at most $n$ CQs, it trivially holds that learning all implementations requires $\mathcal{O}(i(kn^2 + n \log l))$ OQs and $in$ CQs.
~\newline
Now, we consider \textsc{IncrementalFingerprinting}. Fingerprinting one implementation requires at most $m^2$ OQs since the fingerprint contains at most one sequence to separate each pair of models. Fingerprinting all implementations, therefore, requires at most $im^2$ OQs. Under the perfect teacher assumption, we always learn correct models of implementations and only need to learn them if there is no $\Spec \in \Specs$ that is equivalent to the implementation. Therefore, we only need to learn $m = |\Specs|$ models. Thus, the output query complexity is $\mathcal{O}(m(kn^2 + n \log l) + im^2)$. 

When considering the maximum number of CQs, we note that we only learn a model $m$ times, each time requiring at most $n$ CQs. Additionally, the CQ after fingerprinting occurs at most $i$ times. Combining these results, we find that at most $mn + i$ CQs are required.
\end{proof}

\textbf{Analogous Proof for $\mathbf{\adaptivelsharp}$.} First, we note that the complexity of $\adaptivelsharp$ is 
$\mathcal{O}(kn^2 + kno + no^2 + n \log l)$ where $o$ is the number of equivalence classes over all reference models. In our case, $o$ is at most $mn$ as there are $m$ models in $\Specs$ of each at most $n$ states. Therefore, the complexity for $\adaptivelsharp$ is $\mathcal{O}(kn^2 + kmn^2 + n^3m^2 + n \log l)$.

~\newline
Let $m=|\Specs|$, $i=|\Impls|$. Assume that a perfect teacher is available and that all $\M \in \Specs$ have at most $n$ states, $k$ inputs and counterexamples of length at most $l$. Repeated $\adaptivelsharp$ learns the the correct set of models $\Specs$ within $\mathcal{O}(i(kn^2 + kmn^2 + n^3m^2) + mn \log l)$ OQs and at most $mn + i-m$ CQs. \textsc{IncrementalFingerprinting} with $SepSeq$ and $\lsharp$ learns the correct set of models $\Specs$ within $\mathcal{O}(m(kn^2 + kmn^2 + n^3m^2 + n \log l) + im^2)$ OQs and at most $mn + i$ CQs.
\begin{proof}
First, we consider repeated application of $\adaptivelsharp$. Because we have a perfect teacher, we know that we have to learn a model $m$ times and rebuild it $i-m$ times. Rebuilding may take $\mathcal{O}(kn^2 + kmn^2 + n^3m^2)$ OQs but is guaranteed to terminate with all required states (see Thm.~4.8 from~\cite{KrugerJR24StateMatching}) and thus does not require counterexample processing. Therefore, the output query complexity is $\mathcal{O}(i(kn^2 + kmn^2 + n^3m^2) + mn \log l)$. 
For each $\Spec$, we may need up to $n$ CQs and for each $\Impl$ that already has a learnt model in $\Specs$, we only perform a final CQ as rebuilding leads to the first hypothesis being correct. Thus, at most $mn + i - m$ CQs are required.
~\newline
Now, we consider \textsc{IncrementalFingerprinting} with $\adaptivelsharp$. Fingerprinting one implementation requires at most $m^2$ OQs since the fingerprint contains at most one sequence to separate each pair of models. Fingerprinting all implementations, therefore, requires at most $im^2$ OQs. Under the perfect teacher assumption, we always learn correct models of implementations and only need to learn them if there is no $\Spec \in \Specs$ that is equivalent to the implementation. Therefore, we only need to learn $m = |\Specs|$ models. Thus, the output query complexity is $\mathcal{O}(m(kn^2 + kmn^2 + n^3m^2 + n \log l) + im^2)$. 

When considering the maximum number of output queries, we note that we only learn a model $m$ times, each time requiring at most $n$ CQs. Additionally, the CQ after fingerprinting occurs at most $i$ times. Combining these results, we find that at most $mn + i$ CQs are required.
\end{proof}

\textbf{Proof of Theorem 3.}
Let $\Impls$ be a list of implementations and $\Specs_0$ a set of inequivalent models such that $\Impls \subseteq \Specs_0$. Executing \textsc{IncrementalFingerprinting}$_{\mathcal{C}}$ with initial references $\Specs_0$ and implementations $\Impls$ returns $\Specs$ and $\mu$ such that $\Specs_0 = \Specs$ and for $\Impl \in \Impls$, $\mu(\Impl) = \Spec$ iff $\Impl \sim \Spec$ for some $\Spec \in \Specs$.
\begin{proof}
    For each $\Impl \in \Impls$, there must be a $\Spec \in \Specs_0$ such that $\Impl \sim \Spec$ because of assumption  $\Impls \subseteq \Specs_0$. Additionally, we know that \textsc{Fingerprinting} returns at most one model and since there cannot be a separating sequence that shows that $\Impl \nsim \Spec$, the returned model must be $\Spec$. Next, a CQ is performed for $\Impl \sim \Spec$ and this must lead to output \textit{true} as there does not exist a counterexample. Thus, exactly $\Spec \in \Specs_0$ with $\Spec \sim \Impl$ is returned for implementation $\Impl$.
    Additionally, $\Specs = \Specs_0$ because for each $\Impl \in \Impls$ there is a $\Spec \in \Specs_0$ that is equivalent to $\Impl$. Therefore, we never learn a new model that has to be added to $\Specs_0$.
\end{proof}

\subsection{Benchmark Details} \label{sec:model_info}
\begin{table}[H]
    \centering
    \begin{tabular}{lrrrrr}
        \textbf{Model} & \textbf{States} & \textbf{Inputs} & \textbf{Copies} & \textbf{Learn Symbols} & \textbf{EQ Symbols} \\ \hline
        mbedtls/1.1.8/TLS11 & 6 & 11 & 20 & 1065 & 9818 \\
        mbedtls/1.2.7/TLS12 & 6 & 11 & 42 & 1074 & 9323 \\
        mbedtls/1.3.3/TLS12 & 6 & 11 & 18 & 1055 & 12244 \\
        mbedtls/2.1.5/TLS12 & 6 & 11 & 81 & 1028 & 12218 \\
        mbedtls/2.7.8/TLS12 & 6 & 11 & 72 & 603 & 2681 \\
        mbedtls/3.0.0p1/TLS12 & 8 & 11 & 99 & 936 & 5875 \\
        openssl/0.9.7d/TLS10 & 14 & 11 & 5 & 1629 & 1796151 \\
        openssl/0.9.8a/TLS10 & 14 & 11 & 33 & 1284 & 1833540 \\
        openssl/0.9.8l/TLS10 & 10 & 11 & 1 & 801 & 5359 \\
        openssl/0.9.8y/TLS10 & 14 & 11 & 13 & 1560 & 130666 \\
        openssl/0.9.8zh/TLS10 & 11 & 11 & 15 & 1109 & 71365 \\
        openssl/1.0.0g/TLS10 & 11 & 11 & 4 & 993 & 20260 \\
        openssl/1.0.0m/TLS10 & 13 & 11 & 5 & 1306 & 133092 \\
        openssl/1.0.0p/TLS10 & 11 & 11 & 5 & 1065 & 68804 \\
        openssl/1.0.1/TLS10 & 14 & 11 & 11 & 1548 & 167539 \\
        openssl/1.0.1/TLS12 & 13 & 11 & 8 & 1390 & 108994 \\
        openssl/1.0.1d/TLS12 & 13 & 11 & 2 & 1315 & 174942 \\
        openssl/1.0.1r/TLS12 & 11 & 11 & 33 & 1065 & 52090 \\
        openssl/1.0.2d/TLS12 & 10 & 11 & 39 & 932 & 124984 \\
        openssl/1.0.2m/TLS12 & 8 & 11 & 27 & 787 & 127132 \\
        openssl/1.1.0a/TLS12 & 8 & 11 & 39 & 629 & 106845 \\
        openssl/1.1.1g/TLS12 & 8 & 11 & 24 & 614 & 101873 \\
    \end{tabular}
    \caption{Details on TLS models. All models originate form~\cite{janssen2021fingerprinting}. Learned with \lsharp and RandomWp set to stop early when the correct amount of states is reached, averaged over 30 seeds. For more information on the protocols, we refer the reader to \url{https://github.com/Mbed-TLS/mbedtls} and \url{https://github.com/openssl/openssl}.}
    \label{tab:tls_info}
\end{table}

\begin{table}[H]
    \centering
    \begin{tabular}{lrrrrr}
        \textbf{Model} & \textbf{States} & \textbf{Inputs} & \textbf{Copies} & \textbf{Learn Symbols} & \textbf{EQ Symbols} \\ \hline
        DropBearOrig & 17 & 13 & 5 & 3896 & 17705 \\
        DropBear20 & 20 & 13 & 5 & 4208 & 11520 \\
        DropBear22 & 22 & 13 & 5 & 3828 & 13488 \\
        DropBear24 & 24 & 13 & 5 & 4157 & 7604 \\
        DropBear26 & 26 & 13 & 5 & 4523 & 4901 \\
        OpenSSH26 & 26 & 13 & 5 & 5530 & 351107 \\
        OpenSSHOrig & 27 & 13 & 5 & 6294 & 230825 \\
        OpenSSH28 & 28 & 13 & 5 & 5362 & 666117 \\
        OpenSSH29 & 29 & 13 & 5 & 6597 & 719682 \\
        OpenSSH31 & 31 & 13 & 5 & 5911 & 746095 \\
        OpenSSH34 & 34 & 13 & 5 & 6630 & 346615 \\
        OpenSSH36 & 36 & 13 & 5 & 7691 & 688311 \\
        BitVise39 & 39 & 13 & 5 & 10594 & 178497 \\
        BitVise45 & 45 & 13 & 5 & 19569 & 156627 \\
        BitVise47 & 47 & 13 & 5 & 16148 & 143216 \\
        BitVise54 & 54 & 13 & 5 & 23330 & 696630 \\
        BitVise57 & 57 & 13 & 5 & 22720 & 1165158 \\
        BitVise59 & 59 & 13 & 5 & 26440 & 1586899 \\
        BitVise63 & 63 & 13 & 5 & 27342 & 929043 \\
        BitViseOrig & 66 & 13 & 5 & 31236 & 2681762 \\
    \end{tabular}
    \caption{Details on SSH models. Model names ending in `Orig' refer to models from~\cite{VaandragerGRW22} and available on~\url{https://automata.cs.ru.nl/}, the other models are obtained by manually mutating the base models. Learned with \lsharp and RandomWp set to stop early when the correct amount of states is reached, averaged over 30 seeds. }
    \label{tab:ssh_info}
\end{table}

\begin{table}[H]
    \centering
      \begin{tabular}{lrrrrrr}
        \textbf{Model} & \textbf{States} & \textbf{Inputs} & \textbf{Copies} & \textbf{Learn Symbols} & \textbf{EQ Symbols} & \textbf{url} \\
        \hline
        HiveMQ & 7 & 20 & 5 & 2478 & 1000 & \url{https://docs.hivemq.com/}\\
        emqx & 24 & 20 & 5 & 15106 & 5079 & \url{https://www.emqx.com/en}\\
        Mosquitto & 32 & 20 & 5 & 20900 & 7066 & \url{https://mosquitto.org/}\\
        VerneMQ & 19 & 20 & 5 & 10735 & 3222 & \url{https://vernemq.com/}\\
        ejabberd & 53 & 20 & 5 & 67734 & 10217 & \url{https://www.ejabberd.im/}\\
        mochi & 8 & 20 & 5 & 3480 & 1165 & \url{https://github.com/mochi-mqtt/server}
      \end{tabular}
    \caption{Details on MQTT Learning. Learned with \lsharp and StatePrefixOracle with $10$ walks per state and walk length $12$.}
    \label{tab:mqtt_learning}
\end{table}

\begin{table}[H]
    \centering
      \begin{tabular}{lrrrrr}
        \textbf{Model} & \textbf{States} & \textbf{Learn Symbols} & \textbf{EQ Symbols} \\
        \hline
        cc2652r1\_new & 6 &  4531 & 2533 \\
        cyble-416045-02\_new & 2 & 843.3 & 1695.3\\
        explorable & 2 & 654.3 & 1743.3 \\
      \end{tabular}
    \caption{Details on new BLE models. Each model was learned \emph{online} with \lsharp and RandomWp with a maximum of 100 test queries per EQ.}
    \label{tab:new_ble_models}
\end{table}

\begin{table}[H]
    \centering
\begin{tabular}{lrrrrrr}
        \textbf{Model} & \textbf{States} & \textbf{Inputs} & \textbf{Copies} & \textbf{Learn Symbols} & \textbf{EQ Symbols} \\
        \hline
        CYW43455 & 16 & 7 & 5 & 1710 & 941 \\
        cc2650 & 4 & 7 & 5 & 284 & 96 \\
        cc2652r1\_new & 6 & 7 & 5 & 568 & 253 \\
        cc2652r1\_old & 4 & 7 & 5 & 263 & 77 \\
        cyble-416045-02 & 2 & 7 & 5 & 108 & 27 \\
        explorable & 2 & 7 & 5 & 101 & 31 \\
        nRF52832 & 5 & 7 & 5 & 260 & 312 \\
        tesla\_model\_3 & 10 & 7 & 5 & 813 & 456 \\
\end{tabular}
    \caption{Details on BLE Learning from dot models (offline). Learned with \lsharp and RandomWp set to stop early when the correct amount of states is reached, averaged over 30 seeds.}
    \label{tab:ble_learning}
\end{table}

\begin{table}[H]
    \centering
    \begin{tabular}{lrrrrr}
\textbf{Model} & \textbf{States} & \textbf{Inputs} & \textbf{Copies} & \textbf{Learn Symbols} & \textbf{EQ Symbols} \\
\midrule
M1 & 5 & 32 & 5 & 1675 & 833 \\
M2 & 8 & 32 & 5 & 3726 & 6620 \\
M3 & 8 & 32 & 5 & 3730 & 8289 \\
M4 & 7 & 32 & 5 & 3180 & 3062 \\
M5 & 8 & 32 & 5 & 4073 & 11112 \\
M6 & 7 & 32 & 5 & 3188 & 6067 \\
\bottomrule
\end{tabular}
    \caption{Details on BLEDiff Learning from dot models provided by the authors of~\cite{karim2023blediff}, subsequently made input complete and minimized. Learned with \lsharp and RandomWp set to stop early when the correct amount of states is reached, averaged over 30 seeds.}
    \label{tab:blediff_learning}
\end{table}




\subsection{Motivational Experiment Results}
\begin{table}[H]
    \centering
    \begin{adjustbox}{max width=\textwidth}
    \begin{tabular}{ll|rrr|rrrr}
        \toprule
        Algorithm & $|$Initial Models$|$ & Correct Models & Misclassifications & No Matches & Fingerprint Symbols & Conformance Symbols & Learn Symbols & Total Symbols \\
\midrule
$\rlsharp$: RandomWp100 & 0 &  143.7 - 24.1\% & 452.3 - 75.9\% & 0 - 0.0\% 0 & 0 & 0 & 2606357 & 2606357\\
$\rlsharp$: RandomWp500 & 0 &  401.6 - 67.4\% & 194.4 - 32.6\% & 0 - 0.0\% 0 & 0 & 0 & 14266503 & 14266503\\
\midrule
Fingerprint: SepSeq & 11 & 309.4 - 51.9\% & 273.6 - 45.9\% & 13.0 - 2.2\% & 7275 & 0 & 0 & 7275 \\
\ifalg & 11 & 591.7 - 99.3\% & 4.3 - 0.7\% & 0.0 - 0.0\% & 8415 & 3399028 & 100276 & 3507719 \\
\midrule
Fingerprint: SepSeq & 21 & 568.9 - 95.5\% & 27.1 - 4.5\% & 0.0 - 0.0\% & 10388 & 0 & 0 & 10388 \\
\ifalg & 21 & 595.8 - 100.0\% & 0.2 - 0.0\% & 0.0 - 0.0\% & 8270 & 3485626 & 7210 & 3501105 \\
\midrule
Fingerprint: SepSeq & 22 & 596.0 - 100.0\% & 0.0 - 0.0\% & 0.0 - 0.0\% & 10371 & 0 & 0 & 10371 \\
\ifalg & 22 & 596.0 - 100.0\% & 0.0 - 0.0\% & 0.0 - 0.0\% & 8287 & 3491649 & 0 & 3499936 \\
\bottomrule
        \end{tabular}
        \end{adjustbox}
    \caption{Summarized results for the motivational experiment discussed in Section~\ref{sec:motivation}. Conformance symbols are only the symbols used during the CQ directly after fingerprinting, the conformance symbols during learning are included in the learning symbols.}
    \label{tab:motivational_new}
\end{table}

\subsection{Experimental Evaluation Results}
\label{sec:appendix_experimental_evaluation}
\begin{table}[H]
    \centering
    \begin{tabular}{ll|rrrrrr}
        \toprule
        Benchmark & Algorithm & Fingerprint Symbols & Learn Symbols & Total Symbols \\
        \midrule
        BLE & \rlsharp & 0 & 23530 & 23530 \\
        BLE & \radaptivelsharp & 0 & 23612 & 23612 \\
        BLE & \ifalg & 154 & 4843 & 4997 \\
        \midrule
        BLEDiff & \rlsharp & 0 & 127655 & 127655 \\
        BLEDiff & \radaptivelsharp & 0 & 125310 & 125310 \\
        BLEDiff & \ifalg & 169 & 25244 & 25413 \\
        \midrule
        MQTT & \rlsharp & 0 & 494490 & 494490 \\
        MQTT & \radaptivelsharp & 0 & 471005 & 471005 \\
        MQTT & \ifalg & 162 & 101031 & 101193 \\
        \midrule
        SSH & \rlsharp & 0 & 1433586 & 1433586 \\
        SSH & \radaptivelsharp & 0 & 1424811 & 1424811 \\
        SSH & \ifalg & 1476 & 319697 & 321173 \\
        \midrule
        TLS & \rlsharp & 0 & 892491 & 892491 \\
        TLS & \radaptivelsharp & 0 & 957179 & 957179 \\
        TLS & \ifalg & 8760 & 39677 & 48437 \\
        \bottomrule
        \end{tabular}
        \caption{Summarized results for Experiment 1a. }
        \label{tab:exp1a}
\end{table}
        
\begin{table}[H]
    \centering
    \begin{tabular}{lll|r}
        \toprule
        Benchmark & Algorithm & Budget & Correct Models \\
        \midrule
        BLE & \rlsharp & 2500 & 35.6 - 89.0\% \\
        BLE & \radaptivelsharp & 2500 & 39.4 - 98.4\% \\
        BLE & \ifalg & 2500 & 39.2 - 98.1\% \\
        \midrule
        BLEDiff & \rlsharp & 10000 & 17.9 - 59.8\% \\
        BLEDiff & \radaptivelsharp & 10000 & 27.1 - 90.2\% \\
        BLEDiff & \ifalg & 10000 & 27.9 - 92.8\% \\
        \midrule
        MQTT & \rlsharp & 100000 & 26.6 - 88.5\% \\
        MQTT & \radaptivelsharp & 100000 & 29.2 - 97.5\% \\
        MQTT & \ifalg & 100000 & 29.1 - 97.2\% \\
        \midrule
        SSH & \rlsharp & 750000 & 77.2 - 77.2\% \\
        SSH & \radaptivelsharp & 750000 & 91.8 - 91.8\% \\
        SSH & \ifalg & 750000 & 93.3 - 93.3\% \\
        \midrule
        TLS & \rlsharp & 100000 & 465.6 - 78.1\% \\
        TLS & \radaptivelsharp & 100000 & 569.2 - 95.5\% \\
        TLS & \ifalg & 100000 & 584.1 - 98.0\% \\
        \bottomrule
        \end{tabular}
        \caption{Summarized results for Experiment 1b. }
        \label{tab:exp1b}
\end{table}

\begin{table}[H]
    \centering
    \begin{adjustbox}{max width=\textwidth}
\begin{tabular}{ll|r|rrrr}
\toprule
Benchmark & Algorithm & Correct Models & Fingerprinting Symbols & CQ Symbols & Learning Symbols & Total \\
\midrule
BLE & \rlsharp & 39.9 - 99.8\% & 0 & 0 & 155644 & 155644 \\
BLE & \radaptivelsharp & 40.0 - 100.0\% & 0 & 0 & 154085 & 154085 \\
BLE & \ifalg & 40.0 - 100.0\% & 176 & 109314 & 31154 & 140644 \\
\midrule
BLEDiff & \rlsharp & 11.2 - 37.2\% & 0 & 0 & 157645 & 157645 \\
BLEDiff & \radaptivelsharp & 24.9 - 82.8\% & 0 & 0 & 255977 & 255977 \\
BLEDiff & \ifalg & 24.1 - 80.5\% & 265 & 112316 & 67115 & 179695 \\
\midrule
MQTT & \rlsharp & 25.7 - 85.7\% & 0 & 0 & 1221350 & 1221350 \\
MQTT & \radaptivelsharp & 29.8 - 99.3\% & 0 & 0 & 1086956 & 1086956 \\
MQTT & \ifalg & 29.6 - 98.5\% & 171 & 499849 & 263114 & 763134 \\
\midrule
SSH & \rlsharp & 22.9 - 22.9\% & 0 & 0 & 3480605 & 3480605 \\
SSH & \radaptivelsharp & 70.6 - 70.6\% & 0 & 0 & 4780912 & 4780912 \\
SSH & \ifalg & 64.5 - 64.5\% & 1652 & 2496480 & 1579190 & 4077322 \\
\midrule
TLS & \rlsharp & 143.3 - 24.1\% & 0 & 0 & 2595294 & 2595294 \\
TLS & \radaptivelsharp & 546.5 - 91.7\% & 0 & 0 & 4106272 & 4106272 \\
TLS & \ifalg & 545.8 - 91.6\% & 9760 & 3204005 & 198114 & 3411879 \\
\bottomrule
\end{tabular}
\end{adjustbox}
    \caption{Summarized results for Experiment 1c. }
    \label{tab:exp1_add}
\end{table}

\newpage
\begin{table}[H]
    \centering
    \begin{adjustbox}{max width=\textwidth}
\begin{tabular}{ll|r|rrrr}
\toprule
Benchmark & Components & Correct Models & Fingerprint Symbols & CQ Symbols & Learn Symbols & Total Symbols \\
\midrule
BLE & ADG - RandomWord1000 - \radaptivelsharp & 40.0 - 100.0\% & 176 & 672146 & 173516 & 845838 \\
BLE & ADG - RandomWord1000 - \rlsharp & 40.0 - 100.0\% & 176 & 672009 & 173161 & 845346 \\
BLE & ADG - RandomWp100 - \radaptivelsharp & 40.0 - 100.0\% & 176 & 109314 & 31154 & 140644 \\
BLE & ADG - RandomWp100 - \rlsharp & 40.0 - 100.0\% & 176 & 109300 & 30970 & 140446 \\
BLE & ADG - WpK - \radaptivelsharp & 40.0 - 100.0\% & 176 & 1178870 & 294846 & 1473892 \\
BLE & ADG - WpK - \rlsharp & 40.0 - 100.0\% & 176 & 1178870 & 294661 & 1473707 \\
BLE & SepSeq - RandomWord1000 - \radaptivelsharp & 40.0 - 100.0\% & 228 & 672434 & 173226 & 845888 \\
BLE & SepSeq - RandomWord1000 - \rlsharp & 40.0 - 100.0\% & 228 & 671884 & 173239 & 845351 \\
BLE & SepSeq - RandomWp100 - \radaptivelsharp & 40.0 - 100.0\% & 228 & 109292 & 31141 & 140661 \\
BLE & SepSeq - RandomWp100 - \rlsharp & 40.0 - 100.0\% & 228 & 109252 & 30957 & 140437 \\
BLE & SepSeq - WpK - \radaptivelsharp & 40.0 - 100.0\% & 228 & 1178872 & 294833 & 1473933 \\
BLE & SepSeq - WpK - \rlsharp & 40.0 - 100.0\% & 228 & 1178872 & 294568 & 1473668 \\
\midrule
BLEDiff & ADG - RandomWord1000 - \radaptivelsharp & 8.8 - 29.2\% & 263 & 459260 & 217166 & 676689 \\
BLEDiff & ADG - RandomWord1000 - \rlsharp & 5.5 - 18.3\% & 270 & 459092 & 222335 & 681697 \\
BLEDiff & ADG - RandomWp100 - \radaptivelsharp & 24.1 - 80.5\% & 265 & 112316 & 67115 & 179695 \\
BLEDiff & ADG - RandomWp100 - \rlsharp & 20.3 - 67.7\% & 273 & 98970 & 68232 & 167474 \\
BLEDiff & ADG - WpK - \radaptivelsharp & 30.0 - 100.0\% & 203 & 104112844 & 26047499 & 130160546 \\
BLEDiff & ADG - WpK - \rlsharp & 30.0 - 100.0\% & 203 & 104112844 & 26040358 & 130153405 \\
BLEDiff & SepSeq - RandomWord1000 - \radaptivelsharp & 7.2 - 24.0\% & 286 & 472071 & 207137 & 679494 \\
BLEDiff & SepSeq - RandomWord1000 - \rlsharp & 6.0 - 19.8\% & 301 & 474918 & 199665 & 674883 \\
BLEDiff & SepSeq - RandomWp100 - \radaptivelsharp & 26.3 - 87.7\% & 325 & 121525 & 65367 & 187217 \\
BLEDiff & SepSeq - RandomWp100 - \rlsharp & 22.1 - 73.8\% & 294 & 105217 & 65003 & 170514 \\
BLEDiff & SepSeq - WpK - \radaptivelsharp & 30.0 - 100.0\% & 224 & 104112848 & 26047927 & 130160999 \\
BLEDiff & SepSeq - WpK - \rlsharp & 30.0 - 100.0\% & 224 & 104112848 & 26042524 & 130155596 \\
\midrule
MQTT & ADG - RandomWord1000 - \radaptivelsharp & 23.6 - 78.7\% & 267 & 444971 & 423227 & 868466 \\
MQTT & ADG - RandomWord1000 - \rlsharp & 22.0 - 73.3\% & 294 & 423153 & 463513 & 886960 \\
MQTT & ADG - RandomWp100 - \radaptivelsharp & 29.6 - 98.5\% & 171 & 499849 & 263114 & 763134 \\
MQTT & ADG - RandomWp100 - \rlsharp & 29.1 - 97.0\% & 185 & 493004 & 287362 & 780552 \\
MQTT & ADG - WpK - \radaptivelsharp & 30.0 - 100.0\% & 156 & 156252748 & 39304431 & 195557335 \\
MQTT & ADG - WpK - \rlsharp & 30.0 - 100.0\% & 156 & 155831888 & 39179615 & 195011659 \\
MQTT & SepSeq - RandomWord1000 - \radaptivelsharp & 23.5 - 78.3\% & 271 & 447660 & 413195 & 861126 \\
MQTT & SepSeq - RandomWord1000 - \rlsharp & 22.2 - 74.0\% & 297 & 416989 & 455815 & 873102 \\
MQTT & SepSeq - RandomWp100 - \radaptivelsharp & 29.4 - 98.2\% & 170 & 485930 & 286471 & 772571 \\
MQTT & SepSeq - RandomWp100 - \rlsharp & 29.2 - 97.3\% & 175 & 503734 & 262112 & 766022 \\
MQTT & SepSeq - WpK - \radaptivelsharp & 30.0 - 100.0\% & 156 & 156252748 & 39302856 & 195555760 \\
MQTT & SepSeq - WpK - \rlsharp & 30.0 - 100.0\% & 156 & 155845715 & 39174490 & 195020360 \\
\midrule
SSH & ADG - RandomWord1000 - \radaptivelsharp & 22.4 - 22.4\% & 1420 & 1646509 & 802867 & 2450795 \\
SSH & ADG - RandomWord1000 - \rlsharp & 10.8 - 10.8\% & 1671 & 1411471 & 1167240 & 2580381 \\
SSH & ADG - RandomWp100 - \radaptivelsharp & 64.5 - 64.5\% & 1652 & 2496480 & 1579190 & 4077322 \\
SSH & ADG - RandomWp100 - \rlsharp & 37.0 - 37.0\% & 1803 & 2076632 & 1943940 & 4022374 \\
SSH & ADG - WpK - \radaptivelsharp & 100.0 - 100.0\% & 1356 & 202410117 & 51018700 & 253430172 \\
SSH & ADG - WpK - \rlsharp & 100.0 - 100.0\% & 1356 & 203492342 & 51900170 & 255393868 \\
SSH & SepSeq - RandomWord1000 - \radaptivelsharp & 22.0 - 22.0\% & 2189 & 1620057 & 872903 & 2495149 \\
SSH & SepSeq - RandomWord1000 - \rlsharp & 10.8 - 10.8\% & 2511 & 1421740 & 1128794 & 2553044 \\
SSH & SepSeq - RandomWp100 - \radaptivelsharp & 62.0 - 62.0\% & 2546 & 2505740 & 1574980 & 4083266 \\
SSH & SepSeq - RandomWp100 - \rlsharp & 35.8 - 35.8\% & 2589 & 2052354 & 1922384 & 3977326 \\
SSH & SepSeq - WpK - \radaptivelsharp & 100.0 - 100.0\% & 2086 & 202438306 & 51026351 & 253466743 \\
SSH & SepSeq - WpK - \rlsharp & 100.0 - 100.0\% & 2086 & 202608001 & 51790165 & 254400252 \\
\midrule
TLS & ADG - RandomWord1000 - \radaptivelsharp & 353.4 - 59.3\% & 8824 & 12098388 & 561023 & 12668234 \\
TLS & ADG - RandomWord1000 - \rlsharp & 304.9 - 51.1\% & 9808 & 11999406 & 681247 & 12690460 \\
TLS & ADG - RandomWp100 - \radaptivelsharp & 545.8 - 91.6\% & 9760 & 3204005 & 198114 & 3411879 \\
TLS & ADG - RandomWp100 - \rlsharp & 512.2 - 85.9\% & 10689 & 3109059 & 248937 & 3368686 \\
TLS & ADG - WpK - \radaptivelsharp & 558.0 - 93.6\% & 8495 & 146698960 & 5773796 & 152481251 \\
TLS & ADG - WpK - \rlsharp & 558.0 - 93.6\% & 8495 & 146545674 & 5757351 & 152311520 \\
TLS & SepSeq - RandomWord1000 - \radaptivelsharp & 353.4 - 59.3\% & 9905 & 12125882 & 543444 & 12679230 \\
TLS & SepSeq - RandomWord1000 - \rlsharp & 281.8 - 47.3\% & 10893 & 12012321 & 723714 & 12746928 \\
TLS & SepSeq - RandomWp100 - \radaptivelsharp & 527.9 - 88.6\% & 10737 & 3170738 & 197752 & 3379227 \\
TLS & SepSeq - RandomWp100 - \rlsharp & 448.4 - 75.2\% & 11312 & 2977946 & 231426 & 3220684 \\
TLS & SepSeq - WpK - \radaptivelsharp & 558.0 - 93.6\% & 9898 & 146700891 & 5773722 & 152484511 \\
TLS & SepSeq - WpK - \rlsharp & 558.0 - 93.6\% & 9898 & 146547624 & 5750312 & 152307834 \\
\bottomrule
\end{tabular}
\end{adjustbox}
    \caption{Summarized results for Experiment 2.}
    \label{tab:exp2_add}
\end{table}

\begin{table}[]
    \centering
    \begin{adjustbox}{max width=\textwidth}
\begin{tabular}{llrrrrrrrrr}
\toprule
Benchmark & Fingerprinting - Learning & FCQ Wrong & LCQ Wrong & Correct Models & Entering FCQ & End with FCQ & End with LCQ & Fingerprint Symbols & Learn Symbols & Total Symbols \\
\midrule
BLE & RandomWp100 - RandomWp25 & 0.0 - 0.0\% & 0.5 - 1.2\% & 39.5 - 98.8\% & 34 & 32 & 8 & 153 & 12114 & 119862 \\
BLE & RandomWp100 - RandomWp50 & 0.0 - 0.0\% & 0.25 - 0.6\% & 39.8 - 99.4\% & 34 & 32 & 8 & 150 & 18516 & 127475 \\
BLE & RandomWp100 - RandomWp100 & 0.0 - 0.0\% & 0.05 - 0.1\% & 40.0 - 99.9\% & 34 & 32 & 8 & 150 & 31135 & 140270 \\
\midrule
BLEDiff & RandomWp100 - RandomWp25 & 5.0 - 16.7\% & 7.3 - 24.3\% & 17.7 - 59.0\% & 26 & 17 & 13 & 252 & 40607 & 127737 \\
BLEDiff & RandomWp100 - RandomWp50 & 4.3 - 14.3\% & 4.95 - 16.5\% & 20.8 - 69.2\% & 26 & 20 & 10 & 220 & 49448 & 148955 \\
BLEDiff & RandomWp100 - RandomWp100 & 2.7 - 9.0\% & 3.0 - 10.0\% & 24.3 - 81.0\% & 27 & 21 & 9 & 211 & 67635 & 180710 \\
\midrule
MQTT & RandomWp100 - RandomWp25 & 0.05 - 0.2\% & 4.8 - 16.0\% & 25.1 - 83.8\% & 27 & 19 & 11 & 262 & 264869 & 609911 \\
MQTT & RandomWp100 - RandomWp50 & 0.0 - 0.0\% & 1.3 - 4.3\% & 28.7 - 95.7\% & 26 & 23 & 7 & 198 & 234359 & 704225 \\
MQTT & RandomWp100 - RandomWp100 & 0.0 - 0.0\% & 0.35 - 1.2\% & 29.6 - 98.8\% & 27 & 24 & 6 & 171 & 267846 & 767069 \\
\midrule
SSH & RandomWp100 - RandomWp25 & 16.3 - 16.3\% & 27.85 - 27.9\% & 55.9 - 55.9\% & 93 & 58 & 42 & 2004 & 927287 & 2962899 \\
SSH & RandomWp100 - RandomWp50 & 19.9 - 19.9\% & 19.75 - 19.8\% & 60.4 - 60.4\% & 93 & 66 & 34 & 1934 & 1119367 & 3451109 \\
SSH & RandomWp100 - RandomWp100 & 21.3 - 21.3\% & 14.95 - 14.9\% & 63.8 - 63.8\% & 93 & 70 & 30 & 1799 & 1520484 & 4032741 \\
\midrule
TLS & RandomWp100 - RandomWp25 & 54.55 - 9.2\% & 11.8 - 2.0\% & 529.6 - 88.9\% & 589 & 565 & 31 & 10433 & 81330 & 3244653 \\
TLS & RandomWp100 - RandomWp50 & 47.6 - 8.0\% & 8.45 - 1.4\% & 540.0 - 90.6\% & 589 & 568 & 28 & 9813 & 119449 & 3314805 \\
TLS & RandomWp100 - RandomWp100 & 54.2 - 9.1\% & 7.7 - 1.3\% & 534.1 - 89.6\% & 589 & 569 & 27 & 9535 & 198440 & 3383042 \\
\bottomrule
\end{tabular}
\end{adjustbox}
    \caption{Summarized results for Experiment 3.}
    \label{tab:RQ3_impact}
\end{table}

\subsection{Additional Figures and Tables} \label{app:add}
\begin{figure}[H]
    \centering
    \includegraphics[width=0.59\linewidth]{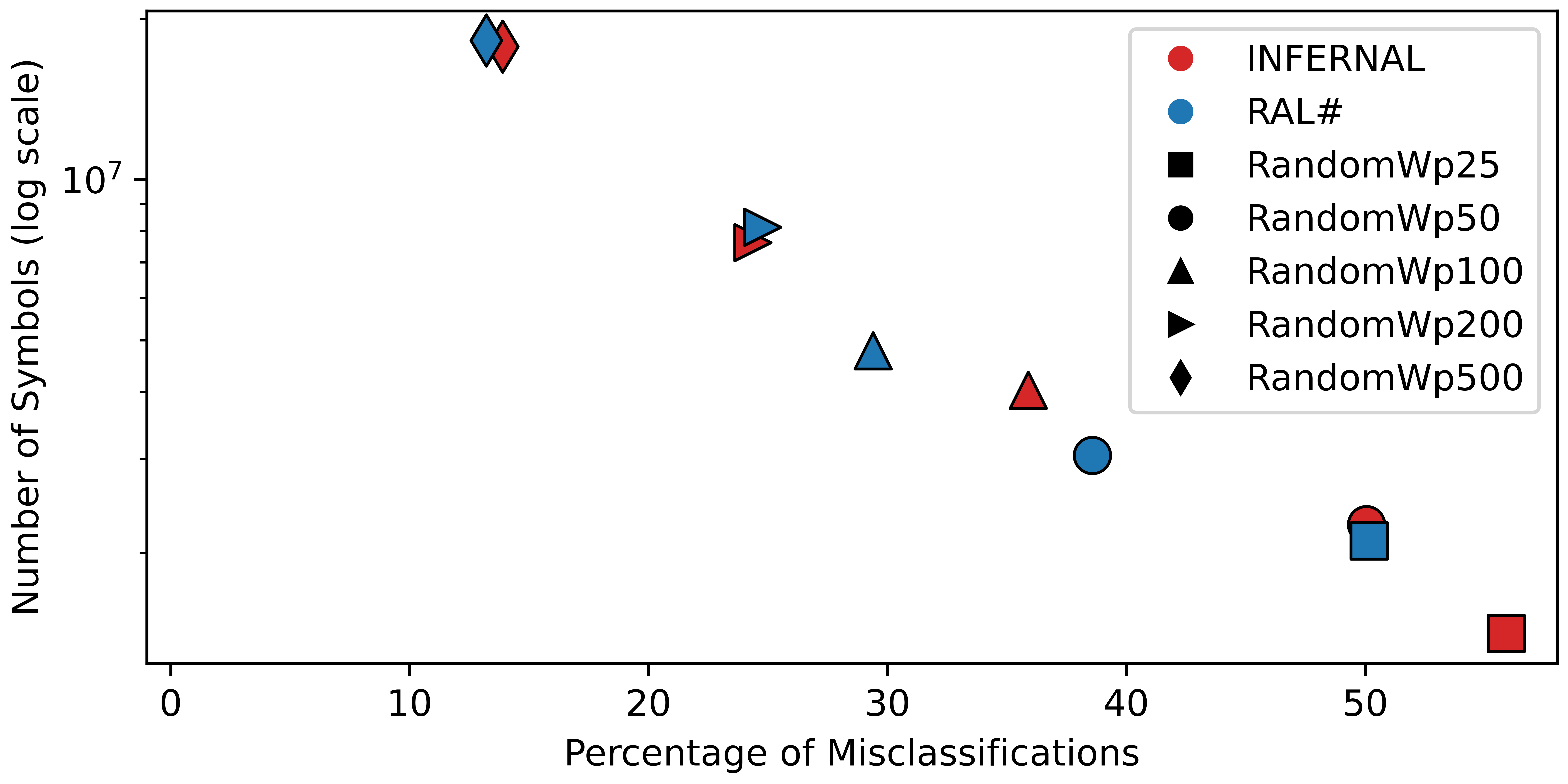}
    \caption{Additional experiment for RQ1. Comparison of \radaptivelsharp and \ifalg for SSH.}
    \label{fig:ALvIFw_SSH}
\end{figure}

\begin{figure}[H]
    \centering
    \includegraphics[width=0.59\linewidth]{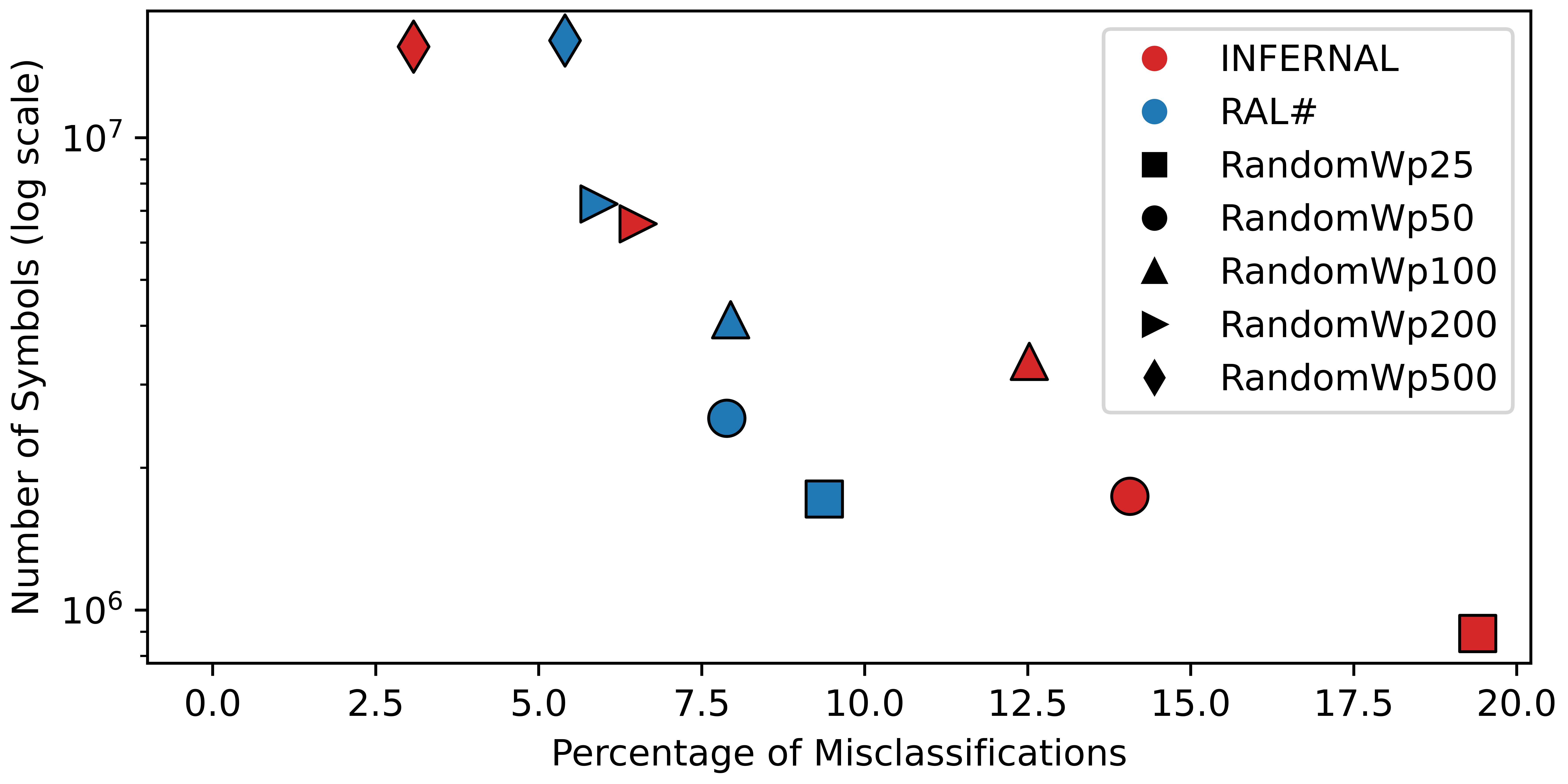}
    \caption{Additional experiment for RQ1. Comparison of \radaptivelsharp and \ifalg for TLS.}
    \label{fig:ALvIFw_TLS}
\end{figure}

\begin{figure}[H]
    \centering
    \includegraphics[width=0.59\linewidth]{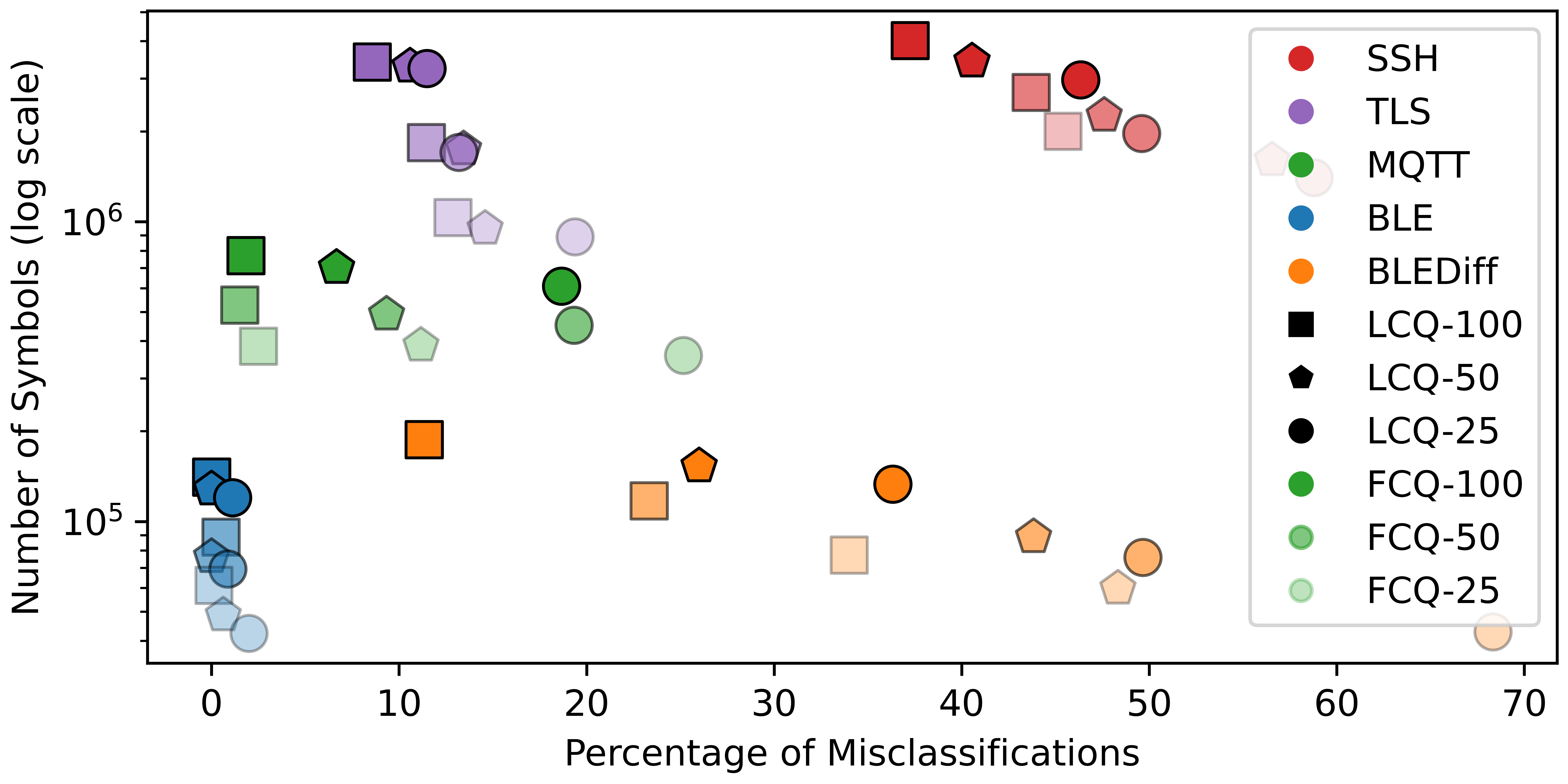}
    \caption{Additional experiment for RQ2. Comparison of the performance of different CQs for fingerprinting and learning.}
    \label{fig:FCQvLCQ}
\end{figure}

\begin{table}[H]
    \centering
    \begin{adjustbox}{max width=\textwidth}
\begin{tabular}{ll|r|rrrr}
\toprule
Benchmark & Fingerprinting - Learning & Correctly Learned  & Fingerprinting & FCQ & Learning & Total \\
\midrule
BLE & RandomWp25 - RandomWp25 & 39.2 - 98.0\% & 140 & 29367 & 12849 & 42356 \\
BLE & RandomWp25 - RandomWp50 & 39.8 - 99.4\% & 133 & 30021 & 18564 & 48717 \\
BLE & RandomWp25 - RandomWp100 & 40.0 - 99.9\% & 134 & 30038 & 31104 & 61276 \\
BLE & RandomWp50 - RandomWp25 & 39.6 - 99.1\% & 137 & 57053 & 12221 & 69410 \\
BLE & RandomWp50 - RandomWp50 & 40.0 - 100.0\% & 134 & 57550 & 18583 & 76267 \\
BLE & RandomWp50 - RandomWp100 & 39.8 - 99.5\% & 133 & 57548 & 31129 & 88810 \\
BLE & RandomWp100 - RandomWp25 & 39.5 - 98.9\% & 148 & 107559 & 12327 & 120033 \\
BLE & RandomWp100 - RandomWp50 & 40.0 - 100.0\% & 134 & 109556 & 18484 & 128174 \\
BLE & RandomWp100 - RandomWp100 & 40.0 - 100.0\% & 134 & 109331 & 31275 & 140739 \\
\midrule
BLEDiff & RandomWp25 - RandomWp25 & 9.5 - 31.7\% & 192 & 17223 & 25423 & 42838 \\
BLEDiff & RandomWp25 - RandomWp50 & 15.5 - 51.7\% & 184 & 21878 & 37776 & 59839 \\
BLEDiff & RandomWp25 - RandomWp100 & 19.8 - 66.0\% & 174 & 26458 & 50697 & 77329 \\
BLEDiff & RandomWp50 - RandomWp25 & 15.1 - 50.3\% & 229 & 42122 & 33543 & 75893 \\
BLEDiff & RandomWp50 - RandomWp50 & 16.9 - 56.2\% & 220 & 43853 & 44783 & 88855 \\
BLEDiff & RandomWp50 - RandomWp100 & 23.0 - 76.7\% & 210 & 56299 & 60793 & 117302 \\
BLEDiff & RandomWp100 - RandomWp25 & 19.1 - 63.7\% & 276 & 90770 & 42153 & 133198 \\
BLEDiff & RandomWp100 - RandomWp50 & 22.2 - 74.0\% & 233 & 103560 & 49281 & 153073 \\
BLEDiff & RandomWp100 - RandomWp100 & 26.6 - 88.7\% & 241 & 121551 & 65867 & 187659 \\
\midrule
MQTT & RandomWp25 - RandomWp25 & 22.4 - 74.8\% & 296 & 82515 & 275074 & 357885 \\
MQTT & RandomWp25 - RandomWp50 & 26.6 - 88.8\% & 213 & 115601 & 270970 & 386784 \\
MQTT & RandomWp25 - RandomWp100 & 29.2 - 97.5\% & 164 & 128308 & 256041 & 384513 \\
MQTT & RandomWp50 - RandomWp25 & 24.2 - 80.7\% & 297 & 174681 & 276447 & 451425 \\
MQTT & RandomWp50 - RandomWp50 & 27.2 - 90.7\% & 223 & 222894 & 267886 & 491003 \\
MQTT & RandomWp50 - RandomWp100 & 29.6 - 98.5\% & 160 & 251157 & 276000 & 527317 \\
MQTT & RandomWp100 - RandomWp25 & 24.4 - 81.3\% & 285 & 322632 & 286519 & 609437 \\
MQTT & RandomWp100 - RandomWp50 & 28.0 - 93.3\% & 215 & 454233 & 248219 & 702667 \\
MQTT & RandomWp100 - RandomWp100 & 29.4 - 98.2\% & 165 & 494615 & 276490 & 771270 \\
\midrule
SSH & RandomWp25 - RandomWp25 & 41.2 - 41.2\% & 1868 & 570473 & 829245 & 1401586 \\
SSH & RandomWp25 - RandomWp50 & 43.5 - 43.5\% & 1792 & 628178 & 975705 & 1605675 \\
SSH & RandomWp25 - RandomWp100 & 54.6 - 54.6\% & 1736 & 667866 & 1333819 & 2003422 \\
SSH & RandomWp50 - RandomWp25 & 50.4 - 50.4\% & 1914 & 1104158 & 862602 & 1968675 \\
SSH & RandomWp50 - RandomWp50 & 52.4 - 52.4\% & 1879 & 1216912 & 1041720 & 2260511 \\
SSH & RandomWp50 - RandomWp100 & 56.3 - 56.3\% & 1770 & 1294165 & 1405565 & 2701500 \\
SSH & RandomWp100 - RandomWp25 & 53.6 - 53.6\% & 2060 & 2062388 & 907471 & 2971920 \\
SSH & RandomWp100 - RandomWp50 & 59.5 - 59.5\% & 1955 & 2360563 & 1066350 & 3428868 \\
SSH & RandomWp100 - RandomWp100 & 62.8 - 62.8\% & 1855 & 2529312 & 1490268 & 4021435 \\
\midrule
TLS & RandomWp25 - RandomWp25 & 480.4 - 80.6\% & 8542 & 799210 & 82231 & 889983 \\
TLS & RandomWp25 - RandomWp50 & 509.1 - 85.4\% & 8337 & 818179 & 122530 & 949046 \\
TLS & RandomWp25 - RandomWp100 & 519.3 - 87.1\% & 8188 & 829912 & 194812 & 1032913 \\
TLS & RandomWp50 - RandomWp25 & 517.4 - 86.8\% & 9083 & 1609942 & 83533 & 1702559 \\
TLS & RandomWp50 - RandomWp50 & 515.9 - 86.6\% & 8739 & 1614072 & 123948 & 1746758 \\
TLS & RandomWp50 - RandomWp100 & 527.8 - 88.5\% & 8667 & 1630764 & 198207 & 1837637 \\
TLS & RandomWp100 - RandomWp25 & 527.5 - 88.5\% & 8760 & 3153798 & 79750 & 3242308 \\
TLS & RandomWp100 - RandomWp50 & 532.9 - 89.4\% & 8940 & 3166015 & 123417 & 3298372 \\
TLS & RandomWp100 - RandomWp100 & 545.0 - 91.4\% & 8570 & 3203933 & 197418 & 3409921 \\
\bottomrule
\end{tabular}
\end{adjustbox}
    \caption{Additional details for Fig.~\ref{fig:FCQvLCQ}.}
    \label{tab:exp3_add}
\end{table}

\begin{table}[h]
    \centering
    \begin{tabular}{ll|r|rrrr}
\toprule
Algorithm & Copies & Correct Models & Fingerprint Symbols & CQ Symbols & Learn Symbols & Total Symbols \\
\midrule
\rlsharp & 0 & 4.8 - 80.0\% & 0 & 0 & 245147 & 245147 \\
\radaptivelsharp & 0 & 5.6 - 93.3\% & 0 & 0 & 233369 & 233369 \\
\ifalg & 0 & 5.6 - 93.3\% & 15 & 89 & 238834 & 238938 \\
\midrule
\rlsharp & 2 & 15.8 - 87.8\% & 0 & 0 & 732097 & 732097 \\
\radaptivelsharp & 2 & 17.4 - 96.7\% & 0 & 0 & 658528 & 658528 \\
\ifalg & 2 & 16.8 - 93.3\% & 101 & 222131 & 308863 & 531096 \\
\midrule
\rlsharp & 4 & 26.6 - 88.7\% & 0 & 0 & 1221398 & 1221398 \\
\radaptivelsharp & 4 & 30.0 - 100.0\% & 0 & 0 & 1092100 & 1092100 \\
\ifalg & 4 & 29.2 - 97.3\% & 215 & 493824 & 291786 & 785824 \\
\midrule
\rlsharp & 6 & 36.6 - 87.1\% & 0 & 0 & 1704670 & 1704670 \\
\radaptivelsharp & 6 & 41.6 - 99.0\% & 0 & 0 & 1516457 & 1516457 \\
\ifalg & 6 & 42.0 - 100.0\% & 243 & 768878 & 238334 & 1007455 \\
\midrule
\rlsharp & 8 & 46.4 - 85.9\% & 0 & 0 & 2173474 & 2173474 \\
\radaptivelsharp & 8 & 53.4 - 98.9\% & 0 & 0 & 1943352 & 1943352 \\
\ifalg & 8 & 53.6 - 99.3\% & 358 & 1023326 & 238891 & 1262575 \\
\midrule
\rlsharp & 10 & 58.0 - 87.9\% & 0 & 0 & 2680479 & 2680479 \\
\radaptivelsharp & 10 & 64.6 - 97.9\% & 0 & 0 & 2367425 & 2367425 \\
\ifalg & 10 & 65.4 - 99.1\% & 492 & 1270763 & 256849 & 1528103 \\
\bottomrule
\end{tabular}
    \caption{Results for MQTT when using a different number of copies per unique model. All algorithms use RandomWp100 and each experiment is repeated 5 times.}
    \label{tab:different_duplicates}
\end{table}

}

\end{document}